\documentclass[12pt]{article}
\pdfoutput=1

\usepackage{cancel,slashed}
\usepackage{bbm}
\usepackage{amssymb}
\usepackage{amsfonts}
\usepackage{amsmath}
\usepackage{graphicx}
\usepackage{cite}

\usepackage{latexsym}
\usepackage{color}
\usepackage{xypic}
\usepackage{cancel,slashed}
\usepackage[hyperfootnotes=false,linktocpage]{hyperref}
\usepackage{color}
\usepackage{transparent}
\usepackage[hang,flushmargin]{footmisc}

\usepackage{blkarray}
\usepackage{multirow}

\newcommand{\bmat}{\left(\begin{array}}
\newcommand{\emat}{\end{array}\right)}

\def\Z{\mathbb{Z}}

\def\CK {{\cal K}}
\def\a {\alpha}

\def\ov{\overline}

\def\IM{\text{Im}\,}
\def\RE{\text{Re}\,}
\def\ov{\overline}
\def\1{{\bf 1}}
\def\2{{\bf 2}}
\def\3{{\bf 3}}
\def\4{{\bf 4}}
\def\6{{\bf 6}}
\def\OR{\Omega\mathcal{R}}

\def\targ#1#2{\genfrac{[}{]}{0pt}{}{#1}{#2}}
\def\targ2#1#2{\genfrac{}{}{0pt}{}{#1}{#2}}

\definecolor{mygr}{rgb}{0,0.6,0}
\definecolor{mygrey}{rgb}{0,0.1,0.2}
\definecolor{myblue}{rgb}{0,0.5,0.9}
\definecolor{myblue2}{rgb}{0,0.5,0.5}
\definecolor{myblue3}{rgb}{0,0.7,0.9}
\definecolor{myblue4}{rgb}{0,0.6,0.6}
\definecolor{myorange}{rgb}{1,0.5,0}
\definecolor{mypurple}{rgb}{0.6,0,1}
\definecolor{mygolden}{rgb}{1,0.8,0.2}
\definecolor{mycyan}{rgb}{0,1,1}
\definecolor{mymagenta}{rgb}{1,0,1}
\definecolor{mykiwi}{rgb}{0.8,1,0.5}
\definecolor{mybrown}{cmyk}{0.14, 0.42, 0.56, 0.2}
\definecolor{myturq}{cmyk}{0.99, 0, 0.2, 0.4}
\definecolor{myaubergine2}{cmyk}{0.4, 0.5, 0, 0.1}
\definecolor{myaubergine}{cmyk}{0.6,0.85,0,0}
\definecolor{CycleGreen}{cmyk}{0.52,0,1,0}
\definecolor{CycleBrown}{cmyk}{0, 0.4, 0.9, 0.2}

\DeclareFontFamily{U}{rcjhbltx}{}
\DeclareFontShape{U}{rcjhbltx}{m}{n}{<->rcjhbltx}{}
\DeclareSymbolFont{hebrewletters}{U}{rcjhbltx}{m}{n}

\DeclareMathSymbol{\lamed}{\mathord}{hebrewletters}{108}
\DeclareMathSymbol{\mem}{\mathord}{hebrewletters}{109}
\DeclareMathSymbol{\ayin}{\mathord}{hebrewletters}{96}
\DeclareMathSymbol{\tsadi}{\mathord}{hebrewletters}{118}
\DeclareMathSymbol{\qof}{\mathord}{hebrewletters}{113}
\DeclareMathSymbol{\resh}{\mathord}{hebrewletters}{114}
\DeclareMathSymbol{\pe}{\mathord}{hebrewletters}{112}
\DeclareMathSymbol{\pesofit}{\mathord}{hebrewletters}{80}
\DeclareMathSymbol{\samekh}{\mathord}{hebrewletters}{115}
\DeclareMathSymbol{\tav}{\mathord}{hebrewletters}{116}
\DeclareMathSymbol{\vav}{\mathord}{hebrewletters}{119}
\DeclareMathSymbol{\het}{\mathord}{hebrewletters}{120}
\DeclareMathSymbol{\yod}{\mathord}{hebrewletters}{121}
\DeclareMathSymbol{\zayin}{\mathord}{hebrewletters}{122}
\DeclareMathSymbol{\alephdot}{\mathord}{hebrewletters}{128}
\DeclareMathSymbol{\tsadisofit}{\mathord}{hebrewletters}{90}
\DeclareMathSymbol{\shin}{\mathord}{hebrewletters}{152}

\makeatletter
\newsavebox\myboxA
\newsavebox\myboxB
\newlength\mylenA

\newcommand*\xoverline[2][0.75]{%
\sbox{\myboxA}{$\m@th#2$}%
\setbox\myboxB\null
\ht\myboxB=\ht\myboxA%
\dp\myboxB=\dp\myboxA%
\wd\myboxB=#1\wd\myboxA
\sbox\myboxB{$\m@th\overline{\copy\myboxB}$}
\setlength\mylenA{\the\wd\myboxA}
\addtolength\mylenA{-\the\wd\myboxB}%
\ifdim\wd\myboxB<\wd\myboxA%
   \rlap{\hskip 0.5\mylenA\usebox\myboxB}{\usebox\myboxA}%
\else
    \hskip -0.5\mylenA\rlap{\usebox\myboxA}{\hskip 0.5\mylenA\usebox\myboxB}%
\fi}
\makeatother

\begin{document}
\pagestyle{plain}

\makeatletter
\@addtoreset{equation}{section}
\makeatother
\renewcommand{\theequation}{\thesection.\arabic{equation}}

\pagestyle{empty}
\rightline{IFT-UAM/CSIC-18-131}
\vspace{0.5cm}
\begin{center}
\Huge{{Type IIA Flux Vacua and $\a'$-corrections}
\\[10mm]}
\large{Dagoberto Escobar, Fernando Marchesano, Wieland Staessens \\[10mm]}
\small{
Instituto de F\'{\i}sica Te\'orica UAM-CSIC, Cantoblanco, 28049 Madrid, Spain
\\[8mm]} 
\small{\bf Abstract} \\[5mm]
\end{center}
\begin{center}
\begin{minipage}[h]{15.0cm} 

We analyse type IIA Calabi-Yau orientifolds with background fluxes, taking into account the effect of perturbative $\alpha'$-corrections. In particular, we consider the $\alpha'$-corrections that modify the metrics in the K\"ahler sector of the compactification. As it has been argued in the literature, including such $\alpha'$-corrections allows to construct the mirror duals of type IIB Calabi-Yau flux compactifications, in which the effect of flux backreaction is under control. We compute the $\alpha'$-corrected scalar potential generated by the presence of RR and NS fluxes, and reformulate it as a bilinear of the flux-axion polynomials invariant under the discrete shift symmetries of the compactification. The use of such invariants allows to express in a compact and simple manner the conditions for Minkowski and AdS flux vacua, and to extract the effect of $\alpha'$-corrections on them.

\end{minipage}
\end{center}
\newpage
\setcounter{page}{1}
\pagestyle{plain}
\renewcommand{\thefootnote}{\arabic{footnote}}
\setcounter{footnote}{0}


\tableofcontents

\section{Introduction}

Compactifications with background fluxes have proven to be a very fertile framework to construct phenomenologically appealing string theory vacua \cite{Douglas:2006es,Denef:2008wq,Ibanez:2012zz,Baumann:2014nda}. A simple and somehow paradigmatic example of such constructions are type IIB orientifold compactifications with three-form fluxes and their F/M-theory counterparts \cite{Dasgupta:1999ss,Becker:1996gj,Gukov:1999ya,Giddings:2001yu}. One important feature of this class of vacua is that one can incorporate background fluxes as quantised harmonic forms on top of an internal Calabi-Yau geometry, and solve for the 10d supergravity equations of motion by simply adding a non-trivial warp factor with a specific internal profile. Given our knowledge of compact Calabi-Yau geometries, this allows to build a plethora of explicit flux compactifications. 

Another interesting feature of this case is that, in the large volume regime, one can easily generate a parametric separation between the Kaluza-Klein scale and the flux-induced mass scale for the Calabi-Yau former moduli. On general grounds, one would expect that when such scale separation is large the modification of the 4d effective theory by the presence of fluxes is minimal, in the sense that they generate a non-trivial superpotential \cite{Gukov:1999ya} but their effect on the K\"ahler metrics is negligible. In this case one may describe the 4d effective potential in terms of a flux-induced superpotential and the K\"ahler potential of the Calabi-Yau compactification, as it is standard practice in most of the flux literature.\footnote{This rule may fail in compactifications with modes localised in regions of strong warping. See e.g. \cite{DeWolfe:2002nn,Giddings:2005ff,Frey:2006wv,Burgess:2006mn,Douglas:2007tu,Shiu:2008ry,Frey:2008xw,Marchesano:2008rg,Martucci:2009sf,Martucci:2014ska,Martucci:2016pzt} for modifications of the effective K\"ahler potential by warping in type IIB compactifications.}

These two attractive features are less established in other corners of the flux landscape, like for instance type IIA flux compactifications. There, it is known that the presence of RR and NS fluxes will modify the internal background geometry to a manifold of $SU(3)$ or $SU(3) \times SU(3)$ structure \cite{Behrndt:2004km,Grana:2004bg,Behrndt:2004mj,Lust:2004ig,Grana:2005sn,Lust:2008zd,Lust:2009zb}, but there is less control on how such a geometry is related to a deformation of a compact Calabi-Yau metric.\footnote{One proposal is to smear out the O6-plane content of a given type IIA flux compactification~\cite{Acharya:2006ne}. However it is not a priori clear~\cite{Blaback:2010sj} if smeared O-planes offer consistent approximate solutions to the string theory equations with localised O-planes.} As a result, a mechanism to generate an abundance of 10d explicit solutions with six compact internal dimensions is lacking, in sharp contrast with the type IIB case. One may address this obstacle by considering Calabi-Yau flux compactifications in the large volume limit, in a regime where fluxes are diluted. In principle, if the 4d cosmological constant and the flux-induced masses of scalars are well below the Kaluza-Klein scale, one should be able to apply the above criterion to argue for the consistency of the flux compactification. Moreover, it should then be possible to implement a systematic search for vacua using a 4d approach, with the flux generated superpotential and the K\"ahler potential of the pure Calabi-Yau case. Nevertheless, since essentially all the compactifications of this kind involve massive type IIA supergravity  \cite{Louis:2002ny,Kachru:2004jr,Grimm:2004ua,DeWolfe:2005uu,Camara:2005dc}, and therefore a Romans' parameter that cannot be diluted as other fluxes, objections have been raised on whether the whole approach is justified \cite{McOrist:2012yc}.

Rather than debating on general grounds, one may instead consider classes of type IIA flux compactifications whose existence is assured by string dualities. For instance, one may consider the mirror duals of the type IIB warped Calabi-Yau compactifications in \cite{Giddings:2001yu}. Just as in the type IIB side one may describe the 4d effective theory in terms of a K\"ahler potential computed from dimensional reduction on a Calabi-Yau, the same should be true in the type IIA side of the mirror map. In other words, for this class of type IIA vacua flux backreaction may take the internal geometry away from a Calabi-Yau metric, but in the same way as the effect of warping can be neglected in many instances, so can the corresponding type IIA deformation. In that case one may safely implement the above 4d approach for a systematic search of vacua, using the naive Calabi-Yau K\"ahler potential and the flux-induced superpotential, as long as the flux-induced masses are well below the compactification scale. 

A fairly general class of type IIA Calabi-Yau compactifications mirror dual to the constructions in \cite{Giddings:2001yu} was analysed in \cite{Palti:2008mg}. As stressed in there, a key ingredient to capture the proper stabilisation of moduli in the type IIB side is the inclusion of (perturbative) $\alpha'$-corrections in the type IIA side, in particular those that affect the K\"ahler moduli sector of the compactification. In fact, together with the fluxes such corrections control the stabilisation of the K\"ahler moduli, and so to some extent determine whether the compactification is in the appropriate regime of validity. 

In this paper we analyse type IIA flux compactifications on Calabi-Yau orientifolds at large or moderately large volume, in the sense that we include the effect of perturbative $\alpha'$-corrections for the K\"ahler sector. We extend the analysis of \cite{Palti:2008mg}, in the sense that we compute the full scalar potential\footnote{More precisely, we consider the classical supergravity potential with perturbative $\alpha'$-corrections, ignoring $g_s$-loop and worldsheet instanton corrections. We expect this to be a good approximation for the compactifications at weak coupling and moderate volumes that we consider. See \cite{Palti:2008mg} for a more detailed discussion of why $g_s$-loop effects should be subdominant. Unlike in \cite{Palti:2008mg} we also neglect the effect of D-brane instantons.} in the presence of general NS and RR fluxes, and apply it to compute both Minkowski and AdS four-dimensional vacua. Our main strategy for this analysis will be to rewrite the Cremmer et al. F-term scalar potential as a bilinear of flux-axion polynomials, namely of the form $V = Z^{AB}\rho_A\rho_B$, as in recent work \cite{Bielleman:2015ina,Carta:2016ynn,Herraez:2018vae,Escobar:2018tiu}. As shown in these references, the classical flux potential can be reformulated as such a bilinear, with $A$ running over the fluxes of the compactification, $\rho_A$ polynomials of the closed and open string axions of the 4d effective theory, and $Z^{AB}$ an (inverse) metric that only depends on their saxionic partners. The polynomial coefficients in the different $\rho_A$ are topological quantities of the compactification, like triple intersection numbers or flux quanta, and such that the $\rho_A$ are invariant under the discrete shift symmetries of the 4d effective theory. As we will show below, this structure is preserved when $\alpha'$-corrections are included with some $\alpha'$-corrections entering the definition of the axion polynomials and others affecting the form of $Z^{AB}$. This shows that the bilinear structure still holds beyond the large volume approximation and, more importantly, in flux compactifications in which the flux backreaction is under control. As in \cite{Herraez:2018vae} using the flux-axion polynomials makes manifest the discrete shift symmetries of the 4d effective theory and, as in \cite{Escobar:2018tiu}, the bilinear formalism allows to implement the search for flux vacua in a more systematic way. Indeed, with our analysis we both recover the results of \cite{Palti:2008mg} and find the $\alpha'$-corrected version of the supersymmetric AdS vacua in \cite{DeWolfe:2005uu,Escobar:2018tiu}.

The paper is organised as follows. In section \ref{S:ClassTypeIIA} we revisit the type IIA flux potential in absence of $\alpha'$-corrections, and its reformulation in terms of a bilinear of axion polynomials reviewing the results of \cite{Herraez:2018vae,Escobar:2018tiu}. In section \ref{S:AlphaCorr} we introduce the effect of perturbative $\alpha'$-corrections in the K\"ahler sector and compute the resulting F-term scalar potential, again rewriting it in terms of $\alpha'$-corrected axion polynomials. With these results, in section \ref{S:AlphaCorrVacua} we compute how $\alpha'$-corrections affect the stabilisation of non-supersymmetric Minkowski and supersymmetric AdS 4d vacua, reproducing previous results in the literature and obtaining new ones. We draw our conclusion in section \ref{sec:con}. Finally, we relegate to Appendix \ref{A:Potential} the technical details regarding the computations of section \ref{S:AlphaCorrVacua}.


\section{The classical type IIA flux potential}\label{S:ClassTypeIIA}

Type IIA flux compactifications offer a unique playground to extract symmetries and structures inherent to (a corner of) the perturbative string landscape. To obtain these landscape properties, the top-down physicist starts from the (tree-level) ten-dimensional type IIA supergravity theory and compactifies it on a suitable background by choice, such as a three-dimensional Calabi-Yau (orientifold) background with internal fluxes. The bottom-up physicist on the other hand will obtain the resulting effective field theory in four dimensions by applying the appropriate supergravity formalism upon specifying the pre-potentials, K\"ahler potentials and/or superpotentials. Irrespective of the chosen approach, it is essential to uncover the special properties of the landscape by using the most suitable formalism. In light of recent results \cite{Bielleman:2015ina,Carta:2016ynn,Herraez:2018vae,Escobar:2018tiu}, it seems that such a formalism could be the reformulation of the scalar potential in terms of shift-invariant axion polynomials, which is the approach taken in this paper. These considerations will be further clarified by this section, which summarises various well-known aspects of type IIA flux compactifications.

 
\subsection{Type IIA flux vacua}\label{Ss:TypeIIAFluxVac}
When compactifying type IIA string theory on a Calabi-Yau three-fold ${\cal CY}_3$, the effective four-dimensional theory is characterised by a residual (local) ${\cal N}=2$ supersymmetry. The Kaluza-Klein (KK) zero-modes of the massless Neveu-Schwarz (NS) and Ramond-Ramond (RR) fields recombine into complex scalar fields (and gauge bosons) filling out the bosonic components of the ${\cal N}=2$ multiplets, i.e. one gravity multiplet, $h^{1,1}({\cal CY}_3)$ vector multiplets, $h^{2,1}({\cal CY}_3)$ hypermultiplets and one tensor multiplet. For a properly defined effective ${\cal N}=2$ supergravity description, the $B_2$-axions are used to complexify the K\"ahler deformations of the Calabi-Yau metric into $h^{1,1}({\cal CY}_3)$ K\"ahler moduli $T^a$:
\begin{equation}\label{Eq:DefKaehlerModuli}
J_c \equiv B + i\, e^{\frac{\phi}{2}} J = T^a \omega_a, \qquad \quad a \in \{1, \ldots, h^{1,1} \}.
\end{equation} 
The massless modes of the NS 2-form and massless K\"ahler deformations associated to the K\"ahler two-form $J$ are in one-to-one with respect to harmonic representatives of the K\"ahler classes $[\ell_s^{-2} \omega_a]$ in $H^{1,1}({\cal M}_6, \Z)$, which are taken to be dimensionless given the insertion of the string length $\ell_s = 2 \pi \sqrt{\alpha'}$. The additional insertion of the ten-dimensional dilaton~$\phi$ indicates that the K\"ahler 2-form is expressed in the Einstein frame. The K\"ahler moduli parameterise the K\"ahler moduli space $\mathfrak{M}_K$ of the Calabi-Yau manifold, which exhibits a K\"ahler structure with K\"ahler potential:
\begin{equation}\label{Eq:KahlerPotKahlerMod}
K_T = - \log \left( \frac{4}{3} \int_{{\cal M}_6} e^{\frac{3\phi}{2}} J \wedge J \wedge J \right) =   - \log \left( \frac{i}{6} {\cal K}_{abc} (T^a - \ov T^a) (T^b - \ov T^b) (T^c - \ov T^c)  \right).
\end{equation} 
The K\"ahler potential depends solely on the internal volume ${\cal V} = \frac{1}{6} \ell_s^{-6} \int_{{\cal M}_6}  J \wedge J \wedge J$, which is expressed as a cubic polynomial in $t^a = \IM(T^a)$ on the righthand side by virtue of (moduli-independent) integral triple intersection numbers ${\cal K}_{abc} = \ell_s^{-6} \int_{{\cal M}_6} \omega_a \wedge \omega_b \wedge \omega_c$. The $B_2$-axions on the other hand do not enter in the K\"ahler potential, which in turn manifests itself in all geometric quantities derived from the K\"ahler potential, such as the moduli space metric.\footnote{Note however that the $b^a$-axions do pop up in the non-canonical couplings between the RR $U(1)$ gauge potentials inherent to the ${\cal N}=2$ vector multiplets. These kinetic and topological mixings between $U(1)$ gauge bosons are equally computed by virtue of the ${\cal N}=2$ pre-potential for the K\"ahler moduli sector.} Furthermore,  the function ${\cal G}_T = e^{- K_T}$ corresponds to a homogenous function of degree three in the geometric  K\"ahler moduli $t^a$, which implies  a no-scale condition for the K\"ahler potential $K_T$:     
\begin{equation}\label{Eq:NoScaleKT}
(K_{T})_{a}(K_T)^{a\ov b} (K_{T})_{\ov b} = 3.
\end{equation}   
The homogeneity of the function ${\cal G}_T$ implies that the K\"ahler moduli space $\mathfrak{M}_K$ is equipped with a special K\"ahler geometry for ${\cal N}=2$ compactifications and that ${\cal G}_T$ is determined in terms of a holomorphic pre-potential ${\cal F}$ by the relation:
\begin{equation}
{\cal G}_T = i\left(  \ov T^A {\cal F}_{T^A}  - T^A  \ov {\cal F}_{ \ov T^A}   \right)_{T^0 = 1},
\end{equation}
where we included a complex coordinate $T^0$ in the set of the K\"ahler moduli $T^A = (T^0, T^a)$ in order to work with homogeneous (projective) coordinates on the K\"ahler moduli space.\footnote{In case one prefers to work in the affine coordinate patch $(1,T^a)$, the relation between the homogenous function ${\cal G}_T$ and the pre-potential has to be properly adjusted:
\begin{equation}
-i \, {\cal G}_T  = 2 {\cal F} - 2 \ov{\cal F} - (T^a - \ov T{}^a) \left( \frac{\partial {\cal F}}{\partial T^a} +  \frac{\partial \ov{\cal F}}{\partial \ov{T}{}^a} \right). 
\end{equation}}
One can then easily check that the K\"ahler potential~\eqref{Eq:KahlerPotKahlerMod} results from the (tree-level) holomorphic pre-potential, valid at large internal volumes: 
\begin{equation}\label{Eq:PrePotTree}
{\cal F}_{\rm tree} (T) = -\frac{1}{3!} \frac{{\cal K}_{abc} T^a T^ bT^c}{T^0}.
\end{equation}
In the next section, we will discuss potential corrections to this pre-potential, which have to be taken into account in regions of the moduli space away from the large volume limit. For now, we shift our focus to the complex structure moduli $z^{\hat \kappa}$ with $\hat \kappa \in \{1,\ldots, h^{2,1}\}$, which reside in the ${\cal N}=2$ hypermultiplets together with the axions emerging as the zero modes of the RR 3-form $C_3$ upon dimensional reduction. The discussion of the complex structure deformations usually starts from a symplectic basis of real harmonic three-forms $(\alpha_\kappa, \beta^\lambda)$ in $H^3({\cal CY}_3)$, in which the Calabi-Yau three-form $\Omega_3$ is expanded:
\begin{equation}
\Omega_3 = Z^\kappa (z) \alpha_\kappa - {\cal F}_\kappa (z) \beta^\kappa, \qquad \kappa \in \{ 0, \ldots, h^{2,1} \}
\end{equation}
with $(Z^\kappa,{\cal F}_\kappa)$ the holomorphic periods depending only on the complex structure moduli~$z^{\hat \kappa}$. Through some algebra, the metric on the complex structure moduli space $\mathfrak{M}_{cs}$, parameterised by the complex structure moduli, can be related to first order derivatives of the holomorphic three-form. Consequently, the moduli space $\mathfrak{M}_{cs}$ is also equipped with a K\"ahler structure through the K\"ahler potential:
\begin{equation}  
K_{cs} = - \log \left( \frac{i}{\ell_s^6} \int_{{\cal CY}_3} \Omega_3 \wedge \ov \Omega_3  \right) = -\log \left( i \ov Z{}^\kappa {\cal F}_\kappa - i Z^\kappa \ov {\cal F}_\kappa \right).
\end{equation}
The expansion of the K\"ahler potential in terms of the holomorphic periods reveals immediately the special K\"ahler property as well, where the periods ${\cal F}_\kappa$ play the role of first order derivatives of a pre-potential ${\cal F}^{cs}$ with respect to the periods $Z^\kappa$. More precisely, the periods ${\cal F}_\kappa$ can be seen as homogeneous functions of degree one in the homogeneous projective coordinates $Z^\kappa$, such that the pre-potential ${\cal F}^{cs} = \frac{1}{2} {Z}^\kappa {\cal F}_\kappa$ is a homogeneous function of degree two. As is well-known, the holomorphic three-form $\Omega_3$ is determined up to a complex phase, which implies the presence of a complex rescaling symmetry $\Omega_3 \rightarrow e^{-h(z)} \Omega_3$ by a holomorphic function $ e^{-h(z)}$. By virtue of this rescaling symmetry, we can set one of the periods to one and work in an affine coordinate patch instead.   

For compactifications on generic Calabi-Yau three-folds with non-vanishing Hodge numbers, the four-dimensional theory exhibits a plethora of massless moduli in the vector multiplets and hypermultiplets. In order to lift these flat directions and break supersymmetry (partially), internal RR- and NS-fluxes can be introduced in the compactification. Adding fluxes along the internal dimensions boils down to adding harmonic components to the exact forms in the RR- and NS-fieldstrength:
\begin{equation} \label{Eq:FieldStrengthsRRNS}
\begin{array}{lcl}
{\bf G} = e^{B_2} \wedge d {\bf A} & \rightarrow & {\bf G} =  e^{B_2} \wedge \left( d {\bf A} + \ov {\bf G} \right),\\
 H_3 = d B_2 & \rightarrow & H_3 = d B_2 + \ov H_3,
\end{array}
\end{equation}
such that the fieldstrengths ${\bf G} = G_0 + G_2 + G_4 + G_6 + G_8 + G_{10}$ and ${H}_3$ still satisfy the Bianchi identities in the absence of localised sources:
\begin{equation}
d\left( e^{-B_2} \wedge  {\bf G}\right) = 0, \qquad dH_3 = 0.
\end{equation}
The Bianchi identities also infer the quantisation of the Page charge (in line with Dirac's philosophy):
\begin{equation}
\frac{1}{\ell_s^{2p-1}} \int_{\pi_{2p}} dA_{2p-1} + \ov G_{2p} \in \Z, \qquad \frac{1}{\ell_s^2} \int_{\pi_3} dB_2 + \ov H_3 \in \Z 
\end{equation}
when integrated over non-trivial homological cycles $\pi_{2p}$ with $p=1,2,3$ for the RR sector and $\pi_3$ for the NS-sector. In the absence of localised sources such as D-branes, the contribution of the exact forms to the Page charge is trivial, such that flux quanta are fully encoded in the harmonic components:
\begin{equation}
\ell_s {\ov G}_0 = m, \qquad \frac{1}{\ell_s} \int_{\tilde \pi^a}{\ov G}_2 = m^a, \qquad \frac{1}{\ell_s^3}\int_{\pi_a} {\ov G}_4 = e_a, \qquad  \frac{1}{\ell_s^5}\int_{{\cal CY}_3} {\ov G}_6 = e_0,
\end{equation}   
with $\tilde \pi^a \in H_2 ({\cal CY}_3,\Z)$ and $\pi_a \in H_4 ({\cal CY}_3, \Z)$. To express the flux quanta associated to the NS 3-form flux,
\begin{equation}
\frac{1}{\ell_s^2} \int_{B^\kappa} {\ov H}_3 = {-} h_\kappa, \qquad  \frac{1}{\ell_s^2}  \int_{A_\lambda} {\ov H}_3 =    h^\lambda,
\end{equation}
we select the set of three-cycles $(A_\lambda,B^\kappa) \in H_3({\cal CY}_3, \Z)$ that are de Rahm duals to the symplectic basis of harmonic three-forms~$(\alpha_\kappa, \beta^\lambda)$. In string theory the flux quanta $\vec{q} = (e_0,e_a,m^a,m,h_\kappa, h^\lambda)$ are integers, while in the low energy supergravity theory these flux parameters are often treated as continuous deformations turning the 4d effective theory into an ${\cal N}=2$ gauged supergravity with masses, St\"uckelberg charges and topological charges. The internal RR-fluxes induce a 4d scalar potential for the K\"ahler moduli whose shape is constrained by a symplectic $Sp(2h^{11} + 2)$ invariance, while the NS-fluxes provide a scalar potential for the complex structure moduli and $C_3$-axions in the hypermultiplets.

The formalism used so far builds heavily on the known geometric properties of Calabi-Yau manifolds, which secretly assume a large internal volume and a weakly coupled dilaton. Away from the large volume regime the background fluxes cannot be considered as diluted and their back-reaction forces us to start from $SU(3)\times SU(3)$ structure manifolds as compactification backgrounds. Nevertheless, one can identify a sub-class of strict $SU(3)$ structure manifolds preserving ${\cal N}=1$ supersymmetry along the four-dimensional directions forming a Minkowski~\cite{Camara:2005dc,Grana:2006kf} or Anti-de Sitter vacua~\cite{Lust:2004ig}. Oftentimes, and in particular to obtain Minkowski flux vacua, localised sources of negative tension such as orientifold planes need to be present in these constructions.   
In this paper we choose to include O6-planes via an orientifold projection that eliminates half of the ${\cal N}=2$ spacetime supersymmetry from the start. More precisely, following \cite{Grimm:2004ua} we consider type IIA Calabi-Yau orientifolds ${\cal M}_6$ that correspond to the quotient manifold of a Calabi-Yau three-fold modded out by the orientifold action $\Omega_p{\cal R} (-)^{F_L}$, with $\Omega_p$ the worldsheet parity operator, $(-)^{F_L}$ the projection operator counting the number of spacetime fermions in the left-moving sector and ${\cal R}$ an anti-holomorphic involution along the internal directions. By looking at the action of the involution on the K\"ahler two-form $J$ and the NS two-form:
\begin{equation}   
{\cal R} (J) = - J, \qquad {\cal R} (B_2) = - B_2,
\end{equation}
one can deduce that the orientifold projection decomposes the $h^{1,1} \; {\cal N}=2$ vector multiplets in the K\"ahler moduli sector to $h^{1,1}_+\; {\cal N}=1$ vector multiplets and $h_-^{1,1}\; {\cal N}=1$ chiral multiplets, with the latter ones containing the K\"ahler deformations of the Calabi-Yau metric. The moduli space for the K\"ahler moduli maintains its special K\"ahler structure, despite the orientifold projection, for which the pre-potential~\eqref{Eq:PrePotTree} now runs over the indices $a\in\{1, \ldots, h_-^{1,1} \}$. The orientifold projection also eliminates part of the RR-fluxes, in line with their transformation properties under the operators $\Omega_p (-)^{F_L}$. The RR two-form flux $\ov G_2$ is only supported on ${\cal R}$-odd two-cycles $\tilde\pi_2^a \in H_2^-({\cal M}_6, \Z)$, while the RR four-form flux $\ov G_4$ only threads ${\cal R}$-even four-cycles $\pi_a \in H_4^+({\cal M}_6, \Z)$. The impact of the orientifold projection on the complex structure moduli sector is more drastic, yet also here the complex structure moduli space does retain its special K\"ahler structure in terms of the redefined complex structure moduli for the ${\cal N}=1$ supersymmetric theory. A proper definition of the ${\cal N}=1$ complex structure moduli starts by considering the action of the anti-holomorphic involution on the Calabi-Yau three-form and the RR three-form $C_3$: 
\begin{equation}
{\cal R} (\Omega_3) = \ov \Omega_3, \qquad {\cal R}(C_3) = C_3.
\end{equation}
Under the anti-holomorphic involution the symplectic basis of three-forms decomposes into a basis of ${\cal R}$-even three-forms $(\alpha_K, \beta^\Lambda) \in H^3_+({\cal M}_6, \Z)$ and ${\cal R}$-odd three-forms $(\beta^K, \alpha_\Lambda) \in H^3_-({\cal M}_6, \Z)$, such that one can easily deduce that the orientifold projection eliminates half of the degree of freedom from the original complex periods in $\Omega_3$. To arrive at the ${\cal N}=1$ complex structure moduli one has to consider instead the complexified three-form $\Omega_c$:
\begin{equation} 
\Omega_c = C_3 + i \, \RE({\cal C} \Omega_3),
\end{equation}
where the compensator field ${\cal C} \equiv e^{-\phi} e^{\frac{1}{2} (K_{cs} - K_T)}$ guarantees the scale-invariance of the holomorphic three-form $\Omega_3 \rightarrow e^{-\RE(h)}\Omega_3$ and the RR three-form insures the holomorphic nature of the complex structure moduli in the orientifolded theory. The independent ${\cal N}=1$ complex structure moduli are thus defined through the complexified three-form by virtue of the ${\cal R}$-odd three-forms:
\begin{equation}\label{Eq:DefComplexStructureModuli}
N^K = \ell_s^{-3} \int_{{\cal M}_6} \Omega_c\wedge \beta^K, \qquad U_\Lambda = \ell_s^{-3} \int_{{\cal M}_6} \Omega_c\wedge \alpha_\Lambda.
\end{equation}
The geometry of the complex structure moduli space $\mathfrak{M}_{cs}$ is characterised by a K\"ahler structure with K\"ahler potential given in terms of the ${\cal N}=1$ complex structure moduli:
\begin{equation}
K_Q = - 2 \log \left( \frac{1}{4} \IM({\cal C} {Z}^\Lambda) \RE({\cal C} {\cal F}_\Lambda) - \frac{1}{4} \RE({\cal C} Z^K) \IM ({\cal C}{\cal F}_K)  \right) = - \log \left( e^{-4D} \right).
\end{equation}
In the last step, we expressed the K\"ahler potential for the complex structure moduli sector in terms of the four-dimensional dilaton $D$ defined through $e^{D} \equiv \frac{e^{\phi}}{\sqrt{ {\cal V}}}$. The periods ${\cal F}_K$ and ${\cal F}_\Lambda$ are not independent, as they correspond to homogeneous functions of degree one in the periods $Z^K$ and $Z^\Lambda$. As such, the function ${\cal G}_Q = e^{-K_Q/2}$ is a homogeneous function of degree two in the complex structure coordinates $n^K = \IM(N^K)$ and $u_\Lambda= \IM (U_\Lambda)$. Hence, the K\"ahler potential $K_Q$ satisfies the following no-scale condition:
\begin{equation} 
(K_Q)_\kappa (K_Q)^{\kappa \ov \lambda} (K_Q)_{\ov \lambda} = 4,
\end{equation}
where $\kappa$, $\lambda$ run over all complex structure moduli $N^K$ and $U_\Lambda$. Similarly to the RR-fluxes, the orientifold projection eliminates part of the NS-fluxes, which are now only supported along the ${\cal R}$-odd three-cycles $(B^K, A_\Lambda)$ de Rahm dual to the three-forms $(\beta^K, \alpha_\Lambda)$.
 
To sum up, both ${\cal N}=2$ and ${\cal N}=1$ Type IIA Calabi-Yau compactifications come with a moduli space that factorises into the product manifold $\mathfrak{M}_K \times \mathfrak{M}_{cs}$, each equipped with a metric obtained from a suitable K\"ahler potential. 
In the presence of (mobile) D6-branes, this innocuous description in terms of a factorised closed string moduli space no longer holds, as the total moduli space in such a set-up also develops directions parameterised by the open string moduli (or D6-brane position moduli), which induce specific mixings in the K\"ahler potential between K\"ahler moduli and complex structure moduli~\cite{Carta:2016ynn,Herraez:2018vae,Escobar:2018tiu}. Upon inclusion of background fluxes, one may wonder if their backreaction may induce mixing as well, as is the case for warped Calabi-Yau type IIB compactifications \cite{Martucci:2009sf}. As in this paper we aim to describe flux vacua in which the flux backreaction can be neglected, we will also neglect their potential effect on the K\"ahler metrics and assume that they only appear in the superpotential, as we now describe. 

\subsection{The bilinear form of the potential}\label{Ss:BilinPot}
The realisation that background closed string fluxes generate a four-dimensional superpotential for the closed string moduli forms a crucial element in the search for string vacua, as the internal fluxes give mass to moduli and stabilise their vacuum expectation value at non-zero values. From a ten-dimensional perspective the background RR-fluxes and NS-fluxes couple to the geometric quantities $J$ and $\Omega_3$ that characterise the internal geometry, such that a four-dimensional superpotential is induced upon integrating out the compact directions \cite{Gukov:1999ya,Taylor:1999ii}:: 
\begin{equation}
W_{\rm flux} = \frac{1}{\ell_s^6} \int_{{\cal M}_6}  {\ov {\bf G}} \wedge e^{J_c} + \Omega_c\wedge H_3 ,
\end{equation}
that is globally well-defined and manifestly gauge-invariant. By virtue of the definitions~\eqref{Eq:DefKaehlerModuli} and~\eqref{Eq:DefComplexStructureModuli} for the closed string moduli and the definitions for the flux quanta, one immediately notices that the superpotential factorises in a purely K\"ahler moduli dependent part:  
\begin{equation} \label{Eq:KahlerSuperpotential}
\ell_s W_T = \int_{{\cal M}_6} {\ov{\bf G}} \wedge e^{ J_c} = e_0 + e_a T^a + \frac{1}{2} {\cal K}_{abc} m^a T^b T^c + \frac{m}{6} {\cal K}_{abc} T^a T^b T^c,
\end{equation}
and a purely complex structure dependent component:
\begin{equation}
\ell_s W_{Q} = \int_{{\cal M}_6} \Omega_c \wedge {\ov H}_3 = h_K N^K  + h^\Lambda U_\Lambda. 
\end{equation}
The structure of these perturbative superpotentials, inherited from ten-dimensional gauge-invariance, allows for a further factorisation in geometric moduli $(t^a,n^K,u_\Lambda)$, axions $(b^a, \xi^K,\xi_\Lambda)$ and a charge vector $\vec{q}$ consisting of the quantised fluxes, i.e.~$\vec{q} = (e_0, e_a, m^a, m, h_K,  h^\Lambda)^t$
\begin{equation}\label{Eq:SuperPotFactForm}
\ell_s(W_T + W_Q) = \vec{\Pi}^t  \cdot (R^{-1})^t \cdot \vec{q}.
\end{equation} 
In this factorisation, the geometric moduli-dependent part is fully captured by the saxion vector $\vec{\Pi}^t (t^a, n^K, u_{\Lambda}) = (1, i t^a, -\frac{1}{2} {\cal K}_{abc} t^b t^c,$ $-\frac{i}{3!} {\cal K}_{abc} t^a t^b t^c, i n^K , i u_{ \Lambda} )$, while a $(2 h_-^{1,1} + h^{2,1} + 3)\times (2 h_-^{1,1} + h^{2,1} + 3)$ dimensional  matrix, 
\begin{equation}
R(b^a,\xi^K, \xi_{\Lambda}) = \left(\begin{array}{cccccc} 
1 & 0 & 0 & 0 &0 & 0 \\
-b^a & \delta^a{}_b & 0 & 0& 0& 0\\
\frac{1}{2} {\cal K}_{abc}  b^b b^c & - {\cal K}_{abc} b^c & \delta^a{}_{b} & 0 & 0 & 0 \\ 
-\frac{1}{3!} {\cal K}_{abc} b^a b^b b^c & \frac{1}{2} {\cal K}_{abc}  b^b b^c & - b^a & 1 & 0& 0  \\
- \xi^K_\star & 0&  0& 0& \delta^K{}_L& 0 \\
-\xi_{\star \Lambda}& 0 & 0& 0& 0& \delta^{\Sigma}{}_\Lambda
 \end{array}\right),
\end{equation}
contains all terms depending on the closed string axions. This rotation matrix is generated through exponentiation 
\begin{equation} \label{Eq:RotMatrixClosedStringAxions}
R^t (b^a, \xi^K, \xi_\Lambda) = e^{b^a P_a + \xi^K P_K + \xi_\Lambda P^\Lambda},
\end{equation}
by a set of nilpotent matrices $P_a$, $P_K$ and $P^\Lambda$: 
\begin{equation}\label{Eq:NilGenShiftCS}
\begin{array}{cc}
\multirow{2}{*}{ $P_a = \left( \begin{array}{cccccc} 0 &  -\vec{\delta}_a^t & 0 & 0& 0 &0\\
0& 0 & -{\cal K}_{abc} & 0 & 0 &0 \\
0& 0& 0 & -\vec{\delta}_a & 0& 0\\
0& 0& 0& 0& 0& 0\\
0& 0& 0& 0& 0& 0\\
0&0&0&0&0& 0
  \end{array}\right),$ }
&  P_K = \left( \begin{array}{cccccc}
  0 & 0& 0 & 0& - \vec{\delta}_K^t  &0\\
0& 0 & 0 & 0 & 0 &0 \\
0& 0& 0 &0 & 0& 0\\
0& 0& 0& 0& 0& 0\\
0& 0& 0& 0& 0& 0\\
    \end{array}\right),\\ 
 & P^\Lambda = \left( \begin{array}{cccccc} 
    0 & 0& 0 & 0& 0&  -(\vec{\delta}^L)^t  \\
0& 0 & 0 & 0 & 0 &0 \\
0& 0& 0 &0 & 0& 0\\
0& 0& 0& 0& 0& 0\\
0& 0& 0& 0& 0& 0\\
  \end{array}\right).
  \end{array} 
\end{equation}
By virtue of these nilpotent generators, the effect of axion shift symmetries on the axion rotation matrix can be deduced in a fairly straightforward way:
\begin{equation}
(R^{-1})^t (b^a +r^a, \xi^K + \varpi^K, \xi_\Lambda+ \varpi_\Lambda ) = (R^{-1})^t (b^a, \xi^K, \xi_\Lambda) e^{-r^a P_a - \varpi^K P_K - \varpi_\Lambda P^\Lambda},
\end{equation}
with $r^a, \varpi^K, \varpi_\Lambda \in \Z$. The invariance of the superpotential~\eqref{Eq:SuperPotFactForm} under the axion shift symmetries is guaranteed provided that the charge vector transforms accordingly,\footnote{Notice that in general \eqref{discshift} will not map a vector of integer entries $\vec{q}$ to a vector of integer entries, as it should for a vector of quantised fluxes. This will be fixed when $\alpha'$-corrections are taken into account.}
\begin{equation}
\vec{q} \quad \rightarrow \quad  e^{r^a P_a + \varpi^K P_K + \varpi_\Lambda P^\Lambda} \vec{q}.
\label{discshift}
\end{equation}
The transformation of the flux quanta under the axion shift symmetries is a property inherent to the multi-branched structure of the vacua for the closed string axions, which is microscopically related to the cancellation of Freed-Witten anomalies for four-dimensional strings in the presence of background fluxes. The latter statement can be discussed in a more explicit way, say for example for the K\"ahler axion $b^a$ associated to the two-form~$\omega_a$, which is Hodge dualised in four dimensions to a two-form. This two-form couples to its respective four-dimensional string arising from an NS5-brane that wraps the Poincar\'e-dual four-cycle ${\rm PD}(\omega_a)$. In case the pull-back $\ov G_{(2p)}\big|_{PD(\omega_a)}$ of a background RR-flux is non-trivial in cohomology, the $b^a$-axionic string develops a Freed-Witten anomaly along the internal directions that needs to be mediated by emitting a $D(6-2p)$-brane wrapping the $(4-2p)$-cycle in the Poincar\'e-dual class of $\ov G_{(2p)}\big|_{PD(\omega_a)}$. With respect to the non-compact directions, this emitted D-brane forms a four-dimensional domain wall bounded by $b^a$-axionic strings. Such a domain wall separates vacua that differ in their 
RR-flux content~\cite{BerasaluceGonzalez:2012zn} and is unstable against the nucleation of holes bounded by axionic strings. This instability is tied to an axion monodromy generated by $P_a$ by which an axion $b^a$ crosses the domain wall and ends up in a vacuum with different RR-fluxes. The transformation of the RR-fluxes under the axion monodromy thus ensures that the entire set-up remains gauge-invariant at all times. Note that these axion shift symmetries form a particular subset of the full gauge transformations under which the field strengths in equation~\eqref{Eq:FieldStrengthsRRNS} remain invariant. Similar considerations hold for the complex structure axions $\xi^K$ and $\xi_\Lambda$, whose respective axionic strings develop Freed-Witten anomalies in the presence of NS-fluxes $\ov H_3$. These Freed-Witten anomalies can be mediated by emitting D2-branes as presented by table~\ref{Tab:StringsDomainWalls}, which summarises all potential Freed-Witten anomalies in the presence of background fluxes and how their mediated by emitting appropriate D-branes. 
\begin{table}[h]
\begin{center}
\hspace*{-0.6in}
\begin{tabular}{|c||c||c||c||c|}
\hline
 \multicolumn{2}{|c||}{\bf String} &{\bf Flux} &  \multicolumn{2}{|c|}{\bf Domain Wall}\\
  \hline
  Axion & Brane Set-up & type &  Brane Set-up & Rank\\
  \hline\hline
$B_2 = b^a \omega_a$ & NS5 on $[\pi_a] \in H_4^+({\cal M}_6,\Z)$& $\ov G_0 = m$ & D6 on $[\pi_a]$ & $m$ \\
$B_2 = b^a \omega_a$ & NS5 on $[\pi_a] \in H_4^+({\cal M}_6,\Z)$& $\ov G_2 = m^a \omega_a$ & D4 on $[{\rm PD}({\ov G_2} \wedge\omega_a)]$ & $\int_{\tilde \pi^a}\omega_c = {\cal K}_{abc} m^b  $ \\
$B_2 = b^a \omega_a$ & NS5 on $[\pi_a] \in H_4^+({\cal M}_6,\Z)$& $\ov G_4 = e_a \tilde \omega^a$ & D2 at point in ${\cal M}_6$  & $\int_{\pi_a} \ov{G}_4 = e_a $ \\
\hline 
$C_3 = \xi^K \alpha_K $ & D4 on $[B^K] \in H_3^-({\cal M}_6, \Z)$  & $\ov H_3 = h_K \beta^K $  & D2 at point in ${\cal M}_6$   & $ \int_{B^K} {\ov H}_3 ={\color{myblue} -} h_K$ \\
$C_3 = - \xi_{\Lambda} \beta^\Lambda $ & D4 on $[A_\Lambda] \in H_3^-({\cal M}_6, \Z)$ & $\ov H_3 = h^\Lambda \alpha_\Lambda $  & D2 at point in ${\cal M}_6$   &  $\int_{A_\Lambda} {\ov H}_3 = {\color{myblue} +} h^\Lambda$ \\
\hline
\end{tabular}
\caption{Summary of 4d axionic strings with their respective attached domain walls arising from Dp- and NS5-branes wrapping internal cycles on a Calabi-Yau manifold with internal flux. \label{Tab:StringsDomainWalls}}
\end{center}
\end{table}

The invariance of the superpotential~\eqref{Eq:SuperPotFactForm} under the axion shift symmetry suggests the existence of a set of gauge-invariant axion polynomials $\ell_s \vec{\rho} \equiv (R^{-1})^t \cdot \vec{q}$, whose explicit component forms are given by,
\begin{equation}\label{Eq:CSgaugeinvariantrho}
\begin{array}{lcl}
\ell_s \rho_0 &=& e_0 + e_a b^a + \frac{1}{2} {\cal K}_{abc} m^a b^b b^c + \frac{m}{6} {\cal K}_{abc} b^a b^b b^c + h_K \xi_\star ^K  + h^\Lambda \xi_{\star \Lambda} , \\
\ell_s  \rho_a &=& e_a + {\cal K}_{abc}  m^b b^c + \frac{m}{2} {\cal K}_{abc} b^b b^c, \\
\ell_s  \tilde \rho^a &=& m^a + m b^a , \\
\ell_s  \tilde \rho &=& m , \\
\ell_s  \hat \rho_K & = & h_K, \\
\ell_s  \hat \rho^\Lambda & = &  h^\Lambda. 
\end{array}
\end{equation}
As such, the above superpotential can be written as the scalar product of the saxion vector $\vec{\Pi}$ with the axion polynomial vector $\vec{\rho}$, or in component-form:
\begin{equation}
W = \rho_0 + i t^a \rho_a  - \frac{1}{2}  {\cal K}_a \tilde \rho^a - \frac{i}{3!}  {\cal K} \tilde \rho + i n^K \hat \rho_K + i u_\Lambda \hat \rho^\Lambda.
\end{equation}
In the large-volume, classical regime in which we are working,\footnote{Which implies neglecting loop, world-sheet and D-brane instanton corrections, as well as modifications of the K\"ahler potential due to the presence of background fluxes.} this factorisation in terms of saxions, axions and flux quanta does not only hold for the ${\cal N}=1$ superpotential, but can be extended to the F-term scalar potential resulting from the background fluxes,
\begin{equation}\label{Eq:ScalarPotFterm}
V_F = \frac{1}{8 \kappa_4^2 \ell_s^2}\, \vec{q}\,{}^t \cdot R^{-1}(b, \xi) \cdot {\cal Z}^{-1}(t, n, u) \cdot R^{-1 t}(b, \xi) \cdot \vec{q},
\end{equation}
where the inverse metric ${\cal Z}^{-1}(t, n, u)$ corresponds to a real, symmetric matrix depending purely on the geometric moduli $(t^a, n^K, u_\Lambda)$:
\begin{equation}\label{Eq:InverseMetricZNoAlpha}
{\cal Z}^{-1} = 8 e^{K} \left(\begin{array}{cccccc} 4 & \\
& K^{a \ov b} \\
&&\frac{4}{9} {\cal K}^2 K_{a \ov b} \\
&&& \frac{1}{9} {\cal K}^2 & \frac{2}{3} {\cal K} n^I &   \frac{2}{3} {\cal K} u_{ \Lambda}\\ 
&&&  \frac{2}{3} {\cal K} n^J & K^{IJ} & K^{I \Sigma} \\
&&& \frac{2}{3} {\cal K} u_{\Sigma}& K^{\Lambda J} & K^{\Lambda \Sigma}
 \end{array} \right).
\end{equation}
Apart from being aesthetically appealing, the formulation in terms of the axion polynomials $\vec{\rho} = (\rho_0, \rho_a, \tilde \rho^a, \tilde \rho, \hat \rho_K, \hat \rho^\Lambda)$ can be used to methodically search for flux vacua in which the axions and (part of the) geometric moduli are stabilised~\cite{Escobar:2018tiu}, in which case the vacuum conditions are written as constraints relating the various axion polynomials to each other. More explicitly, if one is interested in (partly) supersymmetric vacua, one needs to write down the four-dimensional F-terms for the associated ${\cal N}=1$ chiral multiplets in terms of the axion polynomials and find appropriate relations between the latter for the F-terms to vanish. This method turns out to be useful to identify non-supersymmetric Minkowski vacua and supersymmetric Anti-de Sitter vacua, even in the presence of $\alpha'$-corrections as will be discussed in section~\ref{S:AlphaCorrVacua}. Alternatively, one can directly determine the local extrema of the full scalar potential along each axion and geometric moduli. This method is drastically simplified in the axion polynomial language as well, since the first order derivatives of the scalar potential can equally be expressed in terms of the axion polynomials and derivatives of the inverse metric ${\cal Z}^{-1}$ with respect to geometric moduli.





\section{Introducing $\a'$-corrections in Type IIA}\label{S:AlphaCorr}
The previous section provided a short review of the important lessons Type IIA string compactifications on Calabi-Yau (orientifold) with background fluxes have to offer in the large volume limit. If one goes away from regions in the moduli space where the six-dimensional internal volume is huge, quantum corrections such as higher-derivative curvature corrections and worldsheet instanton corrections have to be taken into account. In this section we will investigate how the perturbative $\alpha'$-corrections modify the classical theory by considering how they affect the K\"ahler potential and superpotential in the four-dimensional ${\cal N}=1$ supergravity description. This will in turn also allow us to expose how perturbative $\alpha'$-corrections fit into the axion polynomial formalism and alter the scalar potential.

\subsection{Axion Polynomials  and $\alpha'$-Corrections}
The ${\cal N}=1$ supergravity description of Type IIA orientifold compactifications with K\"ahler potential \eqref{Eq:KahlerPotKahlerMod} is only reliable for sufficiently large internal volumes. Away from this limit, the K\"ahler potential is modified by the so-called $\alpha'$-corrections, which break the no-scale structure of $K_T$ for generic Calabi-Yau manifolds. In the regime of moderately large volumes in which the world-sheet instanton corrections can be neglected, the most relevant $\alpha'$-corrections are those that descend from $(\alpha')^3 R^4$ curvature corrections in the ten-dimensional supergravity action. Following~\cite{Palti:2008mg}, such corrections can be incorporated via a modification of the pre-potential \eqref{Eq:PrePotTree} of the parent ${\cal N}=2$ compactification. In terms of the homogeneous coordinates $T^A=(T^0, T^a)$ in K\"ahler moduli space the most generic (perturbative) pre-potential is given by:
\begin{equation}\label{Eq:alphaCorrPrePot}
{\cal F}_{\rm per} (T)  = - \frac{1}{6} \frac{{\cal K}_{abc} T^a T^b T^c}{T^0} + \frac{1}{2} K^{(1)}_{ab} T^a T^b + K_a^{(2)} T^a T^0  - \frac{i}{2} K^{(3)} (T^0)^2.  
\end{equation}
The first term is the usual tree-level Calabi-Yau volume from (\ref{Eq:KahlerPotKahlerMod}) and the remaining three terms encode different orders of curvature corrections in $\alpha'$. The term proportional to $K^{(3)}$ corresponds to the $(\alpha')^3$-correction and is the only effective contribution to the K\"ahler potential.
In the ${\cal N}=2$ parent compactification, the parameter $K^{(3)} = - \frac{\zeta(3)}{(2\pi)^3}\,  \chi_{{\cal M}_6} \in \mathbb{R}$ is proportional to the Euler characteristic~$\chi_{{\cal M}_6}$ of the compactification manifold ${\cal M}_6$. The corrections $K_{ab}^{(1)}$ and $K_{a}^{(2)}$ correspond respectively to one-loop and two-loop corrections in $\alpha'$, yet do not have a ten-dimensional counterpart due to the lack of a ten-dimensional curvature polynomial with the appropriate features. Their presence can nevertheless be argued from mirror symmetry, which in fact allows to express them in terms of topological quantities of ${\cal M}_6$ like its triple intersection numbers and second Chern class, see e.g. \cite{Grimm:2018cpv,Corvilain:2018lgw}. Their presence is however physically irrelevant at the level of the K\"ahler metrics, as confirmed by their absence in the K\"ahler potential that results from \eqref{Eq:alphaCorrPrePot}: 
\begin{equation}\label{Eq:alphaCorrectedKaehlerPot}
K_T = - \log \left( \frac{4}{3} {\cal K}_{abc} t^a t^b t^c + 2 K^{(3)} \right) = - \log\left(\frac{2}{3} {\cal K} ( 2 + 3 \varepsilon )\right),
\end{equation} 
where the symbol $\varepsilon \equiv \frac{K^{(3)}}{{\cal K}}$ was introduced to capture the $(\alpha')^3$ curvature corrections to the K\"ahler volume ${\cal K} = {\cal K}_{abc} t^a t^b t^c = 6 {\cal V}$. 
As anticipated earlier, in the presence of these perturbative $\alpha'$-corrections the classical no-scale condition for the K\"ahler potential~\eqref{Eq:NoScaleKT} no longer holds and needs to be modified as well:
\begin{equation}
(K_T)_a(K_T)^{a \ov b}  (K_T)_{\ov b} = \frac{3}{1-3\varepsilon}.
\end{equation}
For generic Calabi-Yau compactifications with background fluxes, the (perturbative) $\alpha'$-corrections to the K\"ahler moduli pre-potential (\ref{Eq:alphaCorrPrePot}) also induce corrections to the superpotential for K\"ahler moduli~\cite{Palti:2008mg}. By rewriting the superpotential in terms of the homogeneous coordinates $T^A=(T^0,T^a)$, the $\alpha'$-corrected superpotential can be obtained from the pre-potential (\ref{Eq:alphaCorrPrePot}):
\begin{eqnarray}
\ell_s \left(W_T + W_Q\right) &=& \left( T^0, T^a, -  \partial_{T^a}  {\cal F}_{\rm per}, \partial_ {T^0} {\cal F}_{\rm per}, N^K, U_\Lambda \right)_{T^0 = 1} \cdot \, \vec{q} \; , \end{eqnarray}
with $\vec{q}$ the vector of flux quanta as introduced above. The superpotential $W_Q$ for the complex structure moduli remains unchanged by the curvature corrections, while the part $W_T$ with the K\"ahler moduli takes a similar form as~\eqref{Eq:KahlerSuperpotential}, 
\begin{equation}\label{Eq:SuperpotCorrect}
\ell_s W = \ov e_0 + \ov e_a T^a + \frac{1}{2} {\cal K}_{abc} m^a T^b T^c + \frac{m}{3!} {\cal K}_{abc} T^a T^b T^c -i m K^{(3)}   + h_K N^K + h^\Lambda U_\Lambda
\end{equation}
upon taking into account the curvature correction $K^{(3)}$ and after introducing the curvature corrected flux quanta $\ov e_0 \equiv e_0 - m^a K_a^{(2)}$ and $\ov e_a \equiv e_a - K_{ab}^{(1)} m^b + m K_a^{(2)}$. This clearly shows that the corrections $K_{ab}^{(1)}$ and $K_{a}^{(2)}$ become relevant in the presence of a superpotential for the K\"ahler moduli and cannot be ignored.  
%
%

Let us now see how the inclusion of the curvature corrections is compatible with the axionic shift symmetries of the superpotential. First, notice that its inclusion does not destroy the factorability of the superpotential in terms of geometric moduli and axions. Indeed, we can write the $\alpha'$-corrected flux superpotential as
\begin{equation}\label{Eq:CurvCorrectedSuperpotential}
\ell_s \left(W_T + W_Q\right) = \vec{\ov \Pi}^t \cdot Q  \cdot (\ov R^{-1})^t \cdot \vec{q} ,
\end{equation}
provided that we modify the previous quantities. The saxion vector is now given by $\vec{\ov \Pi}^t (t^a, n^K, u_{\Lambda}) = (1, i t^a, -\frac{1}{2} {\cal K}_{abc} t^b t^c,$ $-\frac{i}{3!} {\cal K}_{abc} t^a t^b t^c - i K^{(3)}, i n^K, i u_{\Lambda} )$, we have introduced a square matrix $Q$ defined below, and the axion rotation matrix is given by
\begin{equation}
\ov R(b^a,\xi^K, \xi_{\Lambda}) = \left(\begin{array}{cccccc} 
1 & 0 & 0 & 0 &0 & 0 \\
-b^a & \delta^a{}_b & 0 & 0& 0& 0\\
\frac{1}{2} {\cal K}_{abc}  b^b b^c +  K_{ba}^{(1)} b^b & - {\cal K}_{abc} b^c & \delta^a{}_{b} & 0 & 0 & 0 \\ 
-\frac{1}{3!} {\cal K}_{abc} b^a b^b b^c - 2 K_a^{(2)} b^a & \frac{1}{2} {\cal K}_{abc}  b^b b^c -  K_{ab}^{(1)} b^b & - b^a & 1 & 0& 0  \\
- \xi^K & 0&  0& 0& \delta^K{}_L& 0 \\ 
-\xi_{\Lambda}& 0 & 0& 0& 0& \delta^{\Sigma}{}_\Lambda
 \end{array}\right).
\end{equation}
Second, the axion rotation matrix is still generated through exponentiation as in~\eqref{Eq:RotMatrixClosedStringAxions}, but now by a modified set of nilpotent, commuting matrices ($\ov P_a, P_K, P^\Lambda)$. The shift-generator $\ov P_a$ for the K\"ahler axions is related to the previous version in~\eqref{Eq:NilGenShiftCS} by conjugation with the charge matrix $Q$,
\begin{equation}\label{Eq:ModifiedPa}
\ov P_a = Q^{-1} P_a Q, \qquad  
Q = \left( \begin{array}{cccccc} 1 & 0 & - K_{a}^{(2)} &0& 0 &0\\
0& \delta^a_b & -K^{(1)}_{ab} & K_{a}^{(2)}  & 0 &0 \\
0& 0& {\delta}_a^b&0 & 0& 0\\
0& 0& 0& 1& 0& 0\\
0& 0& 0& 0& \delta^K_L&  0\\
0&0&0&0&0& \delta^\Sigma_\Lambda
  \end{array}\right).
\end{equation} 
Given these simple extensions, the superpotential remains invariant under the shift symmetries of the closed string axions, provided that the flux quanta transform simultaneously as follows:
\begin{equation}\label{Eq:ChargeTransShift}
\vec{q}\quad \rightarrow \quad  e^{ r^a \ov P_a  + \varpi^K P_K +  \varpi_\Lambda P^\Lambda} \vec{q} \, .
\end{equation}
The transformed flux vector has integer entries provided that $K_{ab}^{(1)}  + \frac{1}{2} \CK_{abb} \in \mathbb{Z}$ and  $2K_a^{(2)} + \frac{1}{6} \CK_{aaa} \in \mathbb{Z}$, $\forall a, b$, which we will assume in the following. Finally, one may express the superpotential in terms of the previous rotation matrix as
\begin{equation}\label{Eq:CurvCorrectedSuperpotentialQ}
\ell_s \left(W_T + W_Q\right) = \vec{\ov \Pi}^t  \cdot (R^{-1})^t \cdot \vec{\ov q}, \qquad \vec{\ov q} \equiv Q \cdot \vec{q}
\end{equation}
Hence, also in the presence of $\alpha'$-corrections one is encouraged to introduce gauge-invariant axion polynomials $\ell_s\vec{\ov\rho} \equiv(R^{-1})^t \cdot \vec{\ov q} $, which can be given explicitly in terms of the flux quanta,
\begin{equation}\label{Eq:CSgaugeinvariantrhoAlphaCor}
\begin{array}{lcl}
\ov \rho_0 &=& \ov e_0 + \ov e_a b^a + \frac{1}{2} {\cal K}_{abc} m^a b^b b^c + \frac{m}{6} {\cal K}_{abc} b^a b^b b^c + h_K \xi_\star ^K  + h^\Lambda \xi_{\star \Lambda}  , \\
\ov{\rho}_a &=& \ov e_a + {\cal K}_{abc}  m^b b^c + \frac{m}{2} {\cal K}_{abc} b^b b^c, \\
{\tilde \rho}^a &=& m^a + m b^a , \\
{\tilde \rho} &=& m , \\
{\hat \rho}_K & = & h_K, \\
{\hat \rho}^\Lambda & = &  h^\Lambda. 
\end{array}
\end{equation}
where $\ov e_0$ and $\ov e_a$ are the curvature-corrected flux quanta as introduced before.


This flux redefinition can be interpreted microscopically by noticing that the curvature corrections $K_{ab}^{(1)}$ and $K_a^{(2)}$ induce lower-dimensional D-brane charge on D-branes wrapping internal cycles, see e.g. \cite{Garcia-Etxebarria:2014wla}. They will, in particular, induce lower-dimensional charge on D-brane domain walls with non-trivial internal worldvolumes. For instance, the $K_a^{(2)}$ curvature corrections induce D4-brane charges on D8-brane domain walls wrapping the full compactification space, and also create bound states of D6-D2 brane domain walls. The $K_{ab}^{(1)}$ curvature corrections on the other hand turn D6-brane domain walls into bound states of D6-D4 brane domain walls. The induced lower-dimensional D-brane charges due to the curvature corrections also imply that the Freed-Witten anomalies associated to the $b$-type axionic strings have to be cured by bound states of domain walls. To end up in a different vacuum separated by the bound states of domain walls, the $b$-axions have to undergo a monodromy transformation generated by the modified matrix $\ov P_a$ defined in~\eqref{Eq:ModifiedPa}. Alternatively, one may redefine the basis of domain walls (or, equivalently the flux basis) by the matrix $Q$ such that the monodromy matrix is generated by $P_a$. The Freed-Witten anomaly cancelation for the $\xi$-type axionic strings on the other hand remains unaffected by the curvature corrections. These considerations offer a microscopic rationale behind the superpotential~\eqref{Eq:CurvCorrectedSuperpotentialQ}, which allows for a factorisation in which curvature corrections of order ${\cal O}(\alpha')$ and ${\cal O}(\alpha'{}^2)$ are incorporated into the set of shift-invariant axion polynomials~\eqref{Eq:CSgaugeinvariantrhoAlphaCor}. The ${\cal O}(\alpha'{}^3)$ curvature contributions represented by $K^{(3)}$ correct the overall volume of the internal space and therefore have to be included in the modified saxion vector $\vec{\ov \Pi}^t$.   

\subsection{The Scalar Potential and $\alpha'$-Corrections}\label{Ss:AlphaCorrScPot}
Since the factorability of the superpotential into saxions and shift-invariant axion polynomials persists in the presence of perturbative $\alpha'$-corrections, one is naturally driven to the question how the modified form of the scalar potential looks like. The most straightforward path to obtain the four-dimensional scalar potential in the presence of background fluxes and perturbative $\alpha'$-corrections consists in computing it directly from the F-term scalar potential:
\begin{equation}\label{Eq:ScPotFterm}   
V_F = \frac{e^{K_T + K_Q}}{\kappa_4^2} \left[\left( \partial_\alpha W + K_\alpha W \right) K^{\alpha \ov \beta} \left( \partial_{\ov \beta} \ov W + K_{\ov \beta} \ov W \right)    - 3 |W|^2\right],
\end{equation}
by inserting the K\"ahler potential~\eqref{Eq:alphaCorrectedKaehlerPot} and superpotential~\eqref{Eq:CurvCorrectedSuperpotentialQ} as obtained in the previous section. In this expression, summation over all closed string moduli is indicated through the Greek letters $(\alpha, \beta)$. 

In practice, the explicit computation of the F-term scalar potential~\eqref{Eq:ScPotFterm} is drastically simplified by deconstructing the expression into three components and applying the elegant formulation of the axion polynomials to the fullest for each component. The first term consists purely of the derivatives of the superpotential with respect to the closed string moduli and requires us to use the modified expressions for the K\"ahler metric as discussed in appendix~\ref{Aa:ModifiedK}:
\begin{equation}\label{Eq:FtermScPotPart1}
\begin{array}{rcl}
\partial_\alpha W K^{\alpha \ov \beta} \partial_{\ov \beta} \ov W& =&   K^{a\ov b} \ov \rho_a \ov \rho_b + \frac{4}{9} {\cal K}^2 (1+\frac{3}{2} \varepsilon)^2 K_{a \ov b} \tilde \rho^a \tilde \rho^b + \frac{1 + 6 \varepsilon}{1 - 3\varepsilon} ( {\cal K}_a \tilde \rho^a)^2  + \frac{1}{3} {\cal K}^2 \tilde \rho^2  \frac{(1+\frac{3}{2} \varepsilon)^2}{1-3\varepsilon}  \\
&& - \frac{4}{3} \tilde \rho {\cal K} \, \ov \rho_a t^a \,  \frac{(1+\frac{3}{2} \varepsilon)^2}{1-3\varepsilon} + K^{N \ov L} \hat \rho_K  \hat \rho_L + K^{N \ov \Lambda} \hat \rho_K  \hat \rho_\Lambda + K^{\Sigma \ov L} \hat \rho_\Sigma  \hat \rho_L + K^{\Sigma \ov \Lambda} \hat \rho_\Sigma  \hat \rho_\Lambda.
\end{array}
\end{equation}
The second component considers the terms without derivatives of the superpotential: 
\begin{equation}\label{Eq:FtermScPotPart2}
\begin{array}{rcl}
K_\alpha K^{\alpha \ov \beta} K_{\ov \beta} |W|^2 - 3 |W|^2 &=& \left(\frac{3}{1-3\varepsilon} + 4 \right) |W|^2 - 3 |W|^2 \\
&=& \frac{4-3\varepsilon}{1-3 \varepsilon} \left[ \left( \RE W \right)^2 + \left( \IM W \right)^2 \right]\\
\end{array}
\end{equation}
where the real and imaginary part of the superpotential can be read off as a function of the axion polynomials directly from the modified superpotential~\eqref{Eq:SuperpotCorrect}. The third and last component consists of the remaining terms containing derivatives of the superpotential, which can be simplified by virtue of relation~\eqref{Eq:ModifiedRelationKabKb} and the holomorphicity of the superpotential:
\begin{equation}\label{Eq:FtermScPotPart3}
\begin{array}{rcl}
K_\alpha W K^{\alpha \ov \beta} \partial_{\ov \beta} \ov W + \partial_\alpha W K^{\alpha \ov \beta} K_{\ov \beta} \ov W &=& -4 \frac{1+ \frac{3}{2} \varepsilon}{1-3\varepsilon} \left( \RE W t^a \partial_{t^a} \RE W + \IM W t^a \partial_{t^a} \IM W \right)\\
&& - 4 \IM W \left(n^K  \hat \rho_K + u_\Lambda \hat \rho^\Lambda  \right) .
\end{array}
\end{equation}
In order to arrive at the simplest expression for the F-term scalar potential further simplifications and manipulations have to be made, which will be discussed at length in appendix~\ref{Aa:ScPotCorrect}. For now, we state the end result of the computation, expressed in terms of the (modified) axion polynomials $(\ov \rho_0, \ov \rho_a, \tilde \rho^a, \tilde \rho, \hat \rho_K, \hat \rho^\Lambda)$:
 \begin{equation}\label{Eq:FullScPotAlpha}
\begin{array}{rcl}
 V_F&=& \frac{ e^{K_T + K_Q}}{\kappa_4^2} \left\{  4 \ov \rho^2_0 + K^{a\ov b} \ov \rho_a \ov \rho_b + \frac{4}{9} {\cal K}^2 \left(1 + \frac{3}{2} \varepsilon\right)^2 K_{cd} \tilde \rho^c\tilde \rho^d + \frac{1}{9} \tilde \rho^2 {\cal K}^2 \left(1+ \frac{3}{2} \varepsilon \right)^2  \right.\\
&&\hspace{0.92in}+ \frac{4}{3} \tilde \rho {\cal K}  \left(1 + \frac{3}{2} \varepsilon\right) ( \hat \rho_K n^K + \hat \rho^\Lambda u_\Lambda ) + K^{K \ov L} \hat \rho_K \hat \rho_L \\
&&\hspace{0.92in} + K^{K \ov \Sigma} \hat \rho_K \hat \rho_\Sigma + K^{\Lambda \ov L} \hat \rho_\Lambda \hat \rho_L  + K^{\Lambda \ov \Sigma} \hat \rho_\Lambda \hat \rho_\Sigma \\
&&\hspace{0.92in}  + \frac{\varepsilon}{1-3\varepsilon} \Big[ 9 \left( \ov \rho_0 + \frac{1}{2}{\cal K}_b \tilde \rho^b   \right)^2 + 9 (n^K \hat \rho_K + \hat \rho^\Lambda u_\Lambda )^2 \\
&&\hspace{1.42in} \left. - 9 \left( t^a \ov \rho_a + \frac{1}{6} \tilde{\rho} {\cal K} (1 -3 \varepsilon) \right)^2  \Big] \right\},
\end{array}
\end{equation}
where now
\begin{equation}
e^{K_T + K_Q} = \frac{e^{4D}}{\frac{4}{3}{\cal K}(1+ \frac{3}{2} \varepsilon)}  \equiv \frac{e^{4\phi}}{8{\cal V}^3(1+ \frac{3}{2} \varepsilon)^3} .
\end{equation}

One notices immediately that the bilinear structure of the F-term scalar potential prevails in the presence of  curvature corrections, such that the scalar potential can still be written as,
\begin{equation}\label{Eq:ScPotAlphaCorrectBil}
V_F =  \frac{1}{8 \kappa_4^2} \, \vec{\ov \rho}\,{}^t \cdot   {\cal Z}^{-1} \cdot \vec{\ov \rho},
\end{equation}
where the inverse metric ${\cal Z}^{-1}$ is now modified by the $K^{(3)}$ curvature corrections expressed in terms of the parameter $\varepsilon$,
\begin{eqnarray}
\label{invZ}
\hspace{-0.6in}{\cal Z}^{-1} &=&8 e^{K_T +K_Q}\left[ \left( 
\begin{array}{cccccc}
4 & 0& 0& 0 & 0& 0 \\ 
0 & K^{a \ov b} & 0& 0 & 0& 0 \\ 
0& 0& \frac{4}{9} {\cal K}^2 K_{a \ov b} (1+\frac{3}{2} \varepsilon)^2 &0 & 0 & 0 \\ 
0& 0&0& \frac{{\cal K}^2}{9}  (1+\frac{3}{2} \varepsilon)^2 & \frac{2}{3} {\cal K} n^L   (1+\frac{3}{2} \varepsilon)   &  \frac{2}{3} {\cal K} u_\Sigma   (1+\frac{3}{2} \varepsilon)  \\ 
0& 0& 0&  \frac{2}{3} {\cal K} n^K   (1+\frac{3}{2} \varepsilon) & K^{K \ov L} & K^{K \ov \Sigma} \\
0& 0& 0&  \frac{2}{3} {\cal K} u_\Lambda   (1+\frac{3}{2} \varepsilon) & K^{\Lambda \ov L} & K^{\Lambda \ov \Sigma} \\
\end{array}
\right)
\right. \notag\\
&&\\
 && \qquad \qquad  \left.  + \frac{\varepsilon}{1 - 3 \varepsilon}\left( \begin{array}{cccccc}
9 & 0& \frac{9}{2} {\cal K}_a & 0 & 0& 0\\
0 & -9 t^a t^b & 0 & -\frac{3}{2} {\cal K} t^a (1-3 \varepsilon) & 0 & 0\\
\frac{9}{2} {\cal K}_b & 0& \frac{9}{4} {\cal K}_a {\cal K}_b & 0& 0 & 0\\
0 & - \frac{3}{2} {\cal K} t^b (1 - 3 \varepsilon)&0&- \frac{{\cal K}^2}{4} (1-3\varepsilon)^2 &0&  0\\
0& 0& 0 &0& 9 n^K n^L & 9 n^K u_\Sigma \\
0& 0& 0 &0 & 9 n^K u_\Lambda & 9 u_\Lambda u_\Sigma \\
\end{array}
\right)
 \right]. \notag
\end{eqnarray}
Due to the curvature corrections, i.e.~$\varepsilon \neq 0$, off-diagonal terms enter in the symmetric matrix. As we will see in the next section, this complicates the search for extrema of the scalar potential at a technical level, but conceptually one may apply the same principles as in \cite{Escobar:2018tiu} to explore the set of vacua in the presence of $\alpha'$-corrections. 


\section{$\a'$-corrected flux vacua}\label{S:AlphaCorrVacua}
The previous section shows how the axion polynomial language allows to incorporate perturbative $\alpha'$-corrections in type IIA Calabi-Yau orientifold compactifications with background fluxes. This insight allowed us to extract the bilinear structure of the scalar potential in terms of the (modified) axion polynomials, but the intricacies of the curvature corrections make the search for vacua of the full perturbative scalar potential quite demanding. This section is therefore devoted to exploiting well-known methods for vacua searches in this context. More precisely, we will extend the results of \cite{Escobar:2018tiu}, that analyses non-supersymmetry Minkowski vacua and supersymmetric Anti-de Sitter vacua in terms of the axion polynomials, to include the effect of curvature corrections. For simplicity, here we will not consider models with mobile D6-branes. 

\subsection{Non-supersymmetric Minkowski Flux Vacua}\label{Ss:ISDVacua}

Following \cite{Palti:2008mg}, one may construct the mirror dual of the no-scale ISD flux vacua of \cite{Giddings:2001yu} by taking a  particular choice of symplectic basis of three-forms with respect to the orientifold projection. For this choice, the complex structure moduli $\{N^K\}_{K\neq 0}$ are projected out and the four-dimensional dilaton $N^0 =  S = \xi^0 + i s$ factorises from the other complex structures moduli $U_\Lambda$ in the K\"ahler potential:
\begin{equation}
\label{KISD}
K_{Q}^{\rm ISD} =  - \log \left( -i (S - \ov S) \right) -2 \log \tilde {\cal G}_Q (u_\Lambda).
\end{equation}
The function $\tilde {\cal G}_Q (u_\Lambda)$ is a homogeneous function of degree $3/2$ of the geometric moduli~$u_\Lambda$. The flux background consists of non-trivial RR-fluxes and a non-vanishing NS-flux $h_0$ supported by the $\OR$-odd three-form $\beta^0$. All other NS-fluxes are set to zero, such that the flux background can be T-dualised to an imaginary self-dual (ISD) three-form $H_3 - S F_3$ in type IIB flux vacua. Due to the absence of the NS-fluxes $h^\Lambda = 0$, the F-terms for the complex structure moduli $U_\Lambda$ reduce to $F_{U_\Lambda} = K_\Lambda W$, and because for these configurations the on-shell superpotential $\langle W \rangle \neq 0$ the  corresponding vacuum configuration breaks supersymmetry spontaneously. Finally, by applying the unbroken no-scale symmetry in the complex structure moduli sector, i.e.~$K_\Lambda K^{\Lambda \ov \Sigma} K_{\ov \Sigma} = 3$, the scalar potential for ISD flux vacua reduces to a positive semi-definite function in terms of the remaining F-terms \cite{Escobar:2018tiu}:
\begin{equation}
\label{ISDsum}
\begin{array}{rcl}
V_F^{\rm ISD} &=& \frac{e^K}{\kappa_4^2} \left[ K^{T^a \ov T^{b}} F_{T^a} F_{\ov T^b} + K^{S\ov S} F_S F_{\ov S} \right]\\
&=& \frac{e^K}{\kappa_4^2}  \left[ 4 \rho_0^2 + K^{a\ov b} \rho_a \rho_b + \frac{4}{9} {\cal K}^2 K_{a\ov b} \tilde \rho^a \tilde \rho^b + \frac{{\cal K}^2}{9} \left(\tilde \rho + \frac{6s}{{\cal K}} \hat \rho_0 \right)^2 \right].
\end{array}
\end{equation}
In the last step we used the bilinear form of the $\alpha'$-uncorrected scalar potential~\eqref{Eq:ScalarPotFterm} to obtain a positive semi-definite expression as a function of the axion polynomials. Based on these expressions for the ISD scalar potential, one immediately sees that the vacuum configuration corresponds to vanishing (uncorrected) F-terms for the dilaton and K\"ahler moduli, or equivalently to the following constraints on the axion polynomials:
\begin{equation}
\rho_0 = 0 = \tilde \rho^a, \quad  \rho_a = 0,  \quad  \tilde \rho + \frac{6s}{{\cal K}} \hat \rho_0 = 0.
\label{ISDFun}
\end{equation}
In this vacuum configuration, the first two constraints stabilise the axion $\xi^0$ and the K\"ahler axions, the third constraint expresses a condition on the flux quanta and the last condition allows for the stabilisation of the four-dimensional dilaton in terms of the overall volume K\"ahler modulus.

In the next phase, we investigate how the ISD flux vacua are modified in the regions of moduli space where the perturbative $\alpha'$-corrections cannot be neglected. As argued in the previous section, the K\"ahler potential for the K\"ahler moduli is modified by the $(\alpha')^3$-correction to expression~\eqref{Eq:alphaCorrectedKaehlerPot}, while the ISD superpotential also requires modifications due to lower order $\alpha'$-corrections. In particular we have that for this case the expression~\eqref{Eq:CurvCorrectedSuperpotentialQ} reduces to
\begin{equation}
\label{WISD}
W_{\rm ISD} = \ov \rho_0 + i \ov \rho_a t^a - \frac{1}{2} {\cal K}_{a} {\tilde \rho}^a - \frac{i}{3!} {\cal K}  {\tilde \rho} - i K^{(3)}  {\tilde \rho} + i s  {\hat \rho}_0.  
\end{equation}
Since the $\alpha'$-corrections do not violate the no-scale symmetry in the $U_\Lambda$-complex structure moduli sector, the first equality of \eqref{ISDsum} still holds, and the same reasoning as above applies to arrive at the vacuum configuration for the ISD flux background. That is, we may derive the Minkowski vacuum conditions by imposing the vanishing of the F-terms for the dilaton and K\"ahler moduli. The dilaton modulus comes with the following F-term in the presence of perturbative $\alpha'$-corrections
\begin{equation}
F_{S}= \frac{1}{2s} \left( i \ov\rho_0  - t^a \ov\rho_a  - \frac{i}{2} {\cal K}_a  {\tilde \rho}^a  +\frac{1}{6} {\cal K} {\tilde \rho}+ K^{(3)}  {\tilde \rho}  + s {\hat \rho}_0  \right),
\label{ISDFS}
\end{equation} 
while the corrected F-term for the K\"ahler moduli reads
\begin{equation}
F_{T^a} = \ov \rho_a  + i {\cal K}_{ab} {\tilde \rho}^a- \frac{1}{2} {\cal K}_a {\tilde \rho}   + \frac{2 i {\cal K}_a}{\frac{4}{3} {\cal K} + 2 K^{(3)}}  W_{\rm ISD} .
\label{ISDFT}
\end{equation}
We may now set both quantities to zero and solve the resulting algebraic equations explicitly. As in \cite{Escobar:2018tiu}, we may simplify such computations by first considering the following linear combination
\begin{equation}
\label{ISDTFT}
\begin{array}{rcl}
t^a F_{T^a} \left( \frac{4}{3} {\cal K} + 2 K^{(3)} \right) &=& 2i {\cal K} \ov\rho_0 + t^a \ov\rho_a (-\frac{2}{3} {\cal K} + 2 K^{(3)}) + i {\cal K}_a {\tilde\rho}^a \left( \frac{1}{3} {\cal K} + 2 K^{(3)} \right)    \\
&&  - {\cal K} {\tilde \rho} \left(  \frac{1}{3} {\cal K} - K^{(3)}\right) - 2 {\hat \rho}_0 s {\cal K}.
\end{array}
\end{equation}
The combined set of the algebraic equations that describe the vacuum constraints for ISD flux vacua can be simply expressed in terms of the redefined axion polynomials~\eqref{Eq:CSgaugeinvariantrhoAlphaCor}. At a first stage one can see that the vanishing of \eqref{ISDFS} and \eqref{ISDTFT} is equivalent to
\begin{equation}
\label{ISDF}
\begin{array}{l@{\hspace{0.4in}}l}
\ov \rho_0 = 0, & - t^a \ov\rho_a+\frac{1}{6} {\cal K} {\tilde \rho} + K^{(3)} {\tilde \rho} + s {\hat \rho}_0 = 0,\\
 {\tilde\rho}^a = 0, & t^a \ov \rho_a (-\frac{2}{3} {\cal K} + 2 K^{(3)} ) + {\cal K} {\tilde \rho} \left( - \frac{1}{3} {\cal K} + K^{(3)} \right) - 2  {\hat \rho}_0 s {\cal K} = 0.
\end{array}
\end{equation}
Notice that the conditions $\ov \rho_0 = 0$ and ${\tilde\rho}^a = 0$ are essentially similar to the uncorrected case \eqref{ISDFun}, while now we no longer have that $\ov\rho_a =0$. The set of equations ${\tilde\rho}^a = 0$ stabilises the K\"ahler axions through the same flux quanta as in absence of $\a'$-corrections, and the axionic partner of the dilaton $\xi^0$ is stabilised by virtue of the condition $\ov{\rho}_0 = 0$, such that its vacuum expectation value can be expressed purely in terms of the curvature corrected flux quanta $\ov e_0$ and $\ov e_a$:
\begin{equation}
h_0 \xi^0 = -\frac{1}{3 m^2} \left( {\cal K}_{abc} m^a m^b m^c  - 3 \ov e_a m^a m\right) - \ov e_0.
\end{equation}
Notice as well that the condition ${\tilde\rho}^a = 0$ and the vanishing eq~\eqref{ISDFT} imply that $\ov \rho_A \propto \CK_a$, and so solving \eqref{ISDF} is equivalent to the vanishing of \eqref{ISDFS} and \eqref{ISDFT}. The remaining two set of equations are solved simultaneously by the relations
\begin{equation}
\label{34ISDcond}
\frac{1}{6} {\cal K} {\tilde \rho} + s {\hat \rho}_0 = \tilde{\rho} K^{(3)}  \frac{\frac{1}{6} \CK + K^{(3)}}{\frac{4}{3}\CK - K^{(3)}}, \quad \quad \ov \rho_a = \ov e_a - \CK_{abc} \frac{m^bm^c}{2m} = \tilde{\rho} K^{(3)}  \frac{\frac{3}{2} \CK_a}{\frac{4}{3}\CK - K^{(3)}},
\end{equation}
which clearly reduce to the previous conditions in the limit $K^{(3)} \rightarrow 0$. They also provide explicit vacuum relations for the dilaton in terms of the flux quanta and curvature corrections:
\begin{equation}
h_0 s = -\frac{1}{6} m {\cal K} + m K^{(3)}  \frac{\frac{1}{6} \CK + K^{(3)}}{\frac{4}{3}\CK - K^{(3)}} = -\frac{1}{6} m \left({\cal K} +6 K^{(3)} \right) + \frac{t^a}{2m} \left( 2m \ov e_a - {\cal K}_{abc} m^b m^c \right),
\end{equation}
as well as for the K\"ahler moduli:
\begin{equation}\label{Eq:KahlerModuliAlphaCorr}
 {\cal K}_a = \frac{(8 {\cal K} - 6 K^{(3)})}{9 m K^{(3)} } \ell_s \ov \rho_a = \frac{(4 {\cal K} - 3 K^{(3)})}{9 m^2 K^{(3)} } \left( 2 \ov e_a m - {\cal K}_{abc} m^b m^c \right).
\end{equation}
in agreement with the results of section 4.2 in \cite{Palti:2008mg}.\footnote{To compare with the results in \cite{Palti:2008mg}, one needs to take into account the difference in conventions for the definition of the fluxes, like a flip in the sign in the Romans' parameter $m$} Finally, one may insert the value of the stabilised moduli into the expression \eqref{WISD} to obtain the on-shell value of the superpotential for this set of vacua:
\begin{equation}
\langle W_{ISD} \rangle  = - \frac{i}{3} \tilde \rho \left({\cal K} +\frac{3}{2} K^{(3)}\right) \frac{{\cal K}-3 K^{(3)}}{{\cal K} - \frac{3}{4} K^{(3)}} .
\end{equation}
As discussed in section 5 of \cite{Escobar:2018tiu} this quantity controls the effective gravitino mass for this set of vacua and, to some extent, the whole spectrum of flux-induced soft-terms in models of intersecting D6-branes. It would be interesting to extract the phenomenological consequences of the $\alpha'$-corrected spectrum of soft-terms in semi-realistic intersecting D6-brane models, a task that we leave for the future. 

From the first equality in \eqref{ISDsum}, that only relies on the choice of K\"ahler metrics (\ref{KISD}) and of NS-fluxes $h^\Lambda =0$, it is clear that the scalar potential is positive semi-definite, as one would expect from the mirror construction in \cite{Giddings:2001yu}. As discussed in \cite{Escobar:2018tiu}, one should be able to see this same feature directly from the bilinear formulation \eqref{Eq:ScPotAlphaCorrectBil} of $V$. Because of the more complicated expression for ${\cal Z}^{-1}$ when $\alpha'$-corrections have been taken into account, showing the positive semi-definiteness of $V$ in this case is more involved. Nevertheless, as we discuss in Appendix \ref{Aa:ISDpot} under the above assumptions one can rewrite \eqref{Eq:ScPotAlphaCorrectBil} as
\begin{equation}\label{Eq:ScPotISDAxPol}
V_F =  \frac{1}{8 \kappa_4^2} \, \vec{\ov \rho}_{\rm ISD}^{\ t} \cdot   {\cal G}^{-1} \cdot \vec{\ov \rho}_{\rm ISD},
\end{equation}
where $\vec{\ov \rho}_{\rm ISD}$ is a shorter vector than $\vec{\ov \rho}$, containing as many entries as RR fluxes, but whose entries are no longer only axion dependent but instead
\begin{equation}
\label{rhoISD}
\vec{\ov \rho}_{\rm ISD}  = \left( \begin{array}{c} \ov \rho_0 \\ \ov \rho_a + \frac{27\varepsilon}{4(1-3\varepsilon) (1+\frac{3}{2}\varepsilon)} \frac{{\cal K}_a}{\CK} s \hat{\rho}_0 \\ \tilde \rho^a \\ \tilde \rho + \frac{6(1- \frac{3}{4}\varepsilon)}{(1-3\varepsilon) (1+\frac{3}{2}\varepsilon)} \frac{s}{\CK} \hat{\rho}_0 \end{array} \right)
\end{equation}
and the symmetric matrix ${\cal G}^{-1}$ is given by
\begin{equation}
\begin{array}{rcl}
{\cal G}^{-1} &=& 8 e^{K_T +K_Q}  \left[ \left( \begin{array}{cccc} 4 & 0 & 0& 0 \\
0 & K^{a\ov b} & 0& 0\\
0& 0& \frac{4}{9}{\cal K}^2 \left( 1+ \frac{3}{2} \varepsilon \right)^2 K_{a\ov b} &0\\
0&0&0& \frac{{\cal K}^2}{9} \left(1+ \frac{3}{2} \varepsilon \right)^2 
 \end{array}\right) \right. + \\
 && \qquad \qquad \left.  + \frac{\varepsilon}{1 - 3 \varepsilon} \left( \begin{array}{cccc} 
 {9} & 0& \frac{9}{2} {\cal K}_a& 0 \\
 0& - {9} t^a t^b &  0 & - \frac{3}{2} {\cal K} t^a (1-3\varepsilon)  \\
 \frac{9}{2} {\cal K}_b  & 0 &  \frac{9}{4} {\cal K}_a {\cal K}_b & 0\\
0& - \frac{3}{2} {\cal K} t^b (1-3\varepsilon) & 0 &  - \frac{{\cal K}^2}{4} (1-3 \varepsilon)^2  \\
  \end{array} \right)  \right].
 \end{array}
\end{equation}
One can easily check that this matrix is positive definite and, in fact, corresponds to the K\"ahler moduli metric derived from the K\"ahler potential \eqref{Eq:alphaCorrectedKaehlerPot}, as a quick comparison with \eqref{invZ} shows. As such, the minima of the potential will only be attained when each of the entries of the vector \eqref{rhoISD} vanish, or in other words upon imposing:
\begin{equation} 
\ov \rho_0 = 0, \quad \tilde \rho^a = 0, \quad \ov \rho_a = - \frac{27\varepsilon}{4(1-3\varepsilon) \left(1+\frac{3}{2}\varepsilon\right)} \frac{{\cal K}_a}{\CK} s \hat{\rho}_0, \quad \tilde \rho = - \frac{6\left(1- \frac{3}{4}\varepsilon\right)}{(1-3\varepsilon) \left(1+\frac{3}{2}\varepsilon\right)} \frac{s}{\CK} \hat{\rho}_0.
\label{ISDcondpsd}
\end{equation}
It is easy to convince oneself that these conditions are equivalent to the relations satisfied in non-supersymmetric Minkowski vacua. Indeed, inserting the last relation in the third one we find that the latter is equivalent to
\begin{equation}
\ov \rho_a = \frac{9 \varepsilon {\cal K}_a}{8\left(1 -\frac{3}{4}\varepsilon\right)} \tilde{\rho} , 
\end{equation}
which is nothing but the second equation in \eqref{34ISDcond}. Similarly, the last relation in \eqref{ISDcondpsd} can be rewritten as
\begin{equation}
\frac{{\cal K}}{6} \tilde \rho + s \hat{\rho}_0  = \tilde \rho \varepsilon \frac{{\cal K}}{8} \frac{1+6\varepsilon}{1-\frac{3}{4}\varepsilon}.
\end{equation}
which is equivalent to the first equation in \eqref{34ISDcond}.

These relations reproduce precisely the proposal of~\cite{Palti:2008mg} to stabilise the K\"ahler moduli by virtue of $\alpha'$-corrections. Whether or not this stabilisation mechanism for the K\"ahler moduli is consistent relies on the possibility of finding a solution to the polynomial equation~\eqref{Eq:KahlerModuliAlphaCorr} for large values of the K\"ahler moduli. In order for the $\alpha'$-correction on the left-hand side to counter-balance the tree-level overall volume, the RR-flux quanta and in particular Roman's mass $m$ have to be chosen appropriately without over-shooting the RR tadpole cancellation conditions. Notice that at the end of this academic exercise, however, the complex structure moduli still remain unstabilised in these non-supersymmetric Minkowski vacua.

\subsection{Supersymmetric AdS vacua}\label{Ss:SUSYVacua}

Just as for non-supersymmetric Minkowski vacua, $\alpha'$-corrections will also affect the conditions that describe AdS supersymmetric vacua in type IIA compactifications. In this case we expect that the effect of $\alpha'$-corrections is a priori less dramatic, in the sense that K\"ahler moduli and complex structure moduli are already stabilised in their absence. Nevertheless, taking into account such corrections may be crucial in setups where moduli are stabilised at moderately large volumes. As we will see in the following, the axion polynomial formalism allows to treat such vacua in a somewhat equal footing as the previous case, and to easily extend the results obtained in \cite{Escobar:2018tiu}, where $\alpha'$-corrections were neglected.

To analyse $\alpha'$-corrected ${\cal N}=1$ AdS vacua we consider a general K\"ahler potential 
\begin{equation}
K = - \log\left( \frac{4}{3} {\cal K} + 2 K^{(3)} \right) - 2 \log\left[ \frac{1}{4} \IM({\cal C} Z^\Lambda) u_\Lambda - \frac{1}{4} n^K \IM ({\cal C} {\cal F}_K) \right]
\end{equation}
and a superpotential given by:
\begin{equation}
W = \ov \rho_0 + i \ov \rho_a t^a - \frac{1}{2} {\cal K}_a \tilde \rho^a - \frac{i}{6} {\cal K} \tilde \rho - i K^{(3)} \tilde \rho + i n^K \hat \rho_K + i u_\Lambda \hat \rho^\Lambda.
\end{equation}
Following the strategy of \cite{Escobar:2018tiu}, we write the different F-terms in terms of axion polynomials and set them to zero:
\begin{equation}\label{Eq:SUSYAdSFterms}
\begin{array}{rcl}
F_{N^K}& =&  \hat \rho_K -   i \frac{\IM({\cal C F}_K)}{2{\cal G}_Q} \left( W_T + W_Q\right) = 0,  \\
F_{U_\Lambda}&=&  \hat \rho^\Lambda +   i \frac{ \IM({\cal C Z}^\Lambda)}{2{\cal G}_Q} \left( W_T + W_Q\right) = 0,\\
F_{T^a} & =& \ov\rho_a +i {\cal K}_{ab} \tilde \rho^b - \frac{1}{2} {\cal K}_a \tilde\rho  +\frac{2i\,{\cal K}_a}{\frac{4}{3} {\cal K} + 2 K^{(3)}}  \left( W_T + W_Q\right)=0.
\end{array}
\end{equation}
Analogously to our previous discussion, simpler equations are obtained when we consider certain linear combinations of  complex structure F-terms
\begin{equation}
\sum_{K=0}^{h^{}} n^K_\star F_{N^K_\star} + \sum_{\Lambda=0}^{h^{}} u_{\star \Lambda} F_{U_{\star \Lambda}} = \sum_{K=0}^{h^{}} \hat \rho _K n^K_\star +  \sum_{\Lambda=0}^{h^{}} \hat \rho^\Lambda u_{\star \Lambda}  +2i \left( W_T + W_Q\right) = 0,
\end{equation}
from where we find the following relations:
\begin{equation}\label{Eq:SUSYAdS1}
\ov \rho_0 - \frac{1}{2} {\cal K}_a \tilde \rho^a = 0, \qquad n^K_\star \hat \rho_K + u_{\star\Lambda} \hat \rho^\Lambda  = \frac{1}{3} {\cal K} \tilde \rho - 2 t^a \rho_a + 2 K^{(3)} \tilde \rho.
\end{equation}
The same can be done with the K\"ahler moduli F-terms, obtaining:
\begin{equation}
\left( \frac{4}{3} {\cal K} + K^{(3)} \right)t^a F_{T^a}  =  t^a \ov \rho_a  \left( \frac{10}{3} {\cal K} + 2 K^{(3)} \right) + i {\cal K}_a \tilde \rho^a \left( \frac{4}{3} {\cal K} + 2 K^{(3)}  \right) - {\cal K} \tilde \rho \left( {\cal K} + 3 K^{(3)}  \right),
\label{sumKahler}
\end{equation}
were we have used \eqref{Eq:SUSYAdS1} to simplify the rhs. It is easy to see that this last one can vanish if $\tilde{\rho}^a=0$, which in turn implies that  $\ov \rho_a \propto \CK_a$ and the vanishing of \eqref{sumKahler} is the only non-trivial F-term condition in the K\"ahler sector. Combining such a condition with the first equation in \eqref{Eq:SUSYAdS1} one obtains the following vacuum relations
\begin{equation}
\ov \rho_0 = 0 , \qquad \tilde \rho^a = 0, \qquad \ov \rho_a =\frac{3}{10}  {\tilde \rho}\, {\cal K}_{a} \left[  \frac{\CK + 3K^{(3)}}{ \CK + \frac{3}{5} K^{(3)}}\right], 
\label{3condads}
\end{equation}
which generalise the conditions obtained in \cite{Escobar:2018tiu}. Comparing to eq.~(3.36) therein, only the third condition is essentially different from the uncorrected case. On the one hand, since the first two conditions are the ones that implement the stabilisation of K\"ahler axions and one linear combination of complex structure moduli, their vacuum expectation values in terms of the fluxes will have a similar form as in  \cite{Escobar:2018tiu}
\begin{equation}\label{axionstab}
h_K \xi_\star^K + h^\Lambda \xi_{\star\, \Lambda} = -\frac{\ov e_0 m^2 - m \ov e_a m^a + \frac{1}{3} {\cal K}_{abc} m^a m^b m^c}{m^2}, \qquad b^a = - \frac{m^a}{m}.
\end{equation}
On the other hand, the geometric part of the K\"ahler moduli, which are stabilised in terms of the background fluxes by the third condition in \eqref{3condads}, will be affected nontrivially by the $(\alpha')^3$-correction term $K^{(3)}$. 

To proceed, we may insert these conditions and the second equation in \eqref{Eq:SUSYAdS1} to obtain the vacuum expectation value for the superpotential in these AdS vacua, finding that
\begin{equation}
\langle W_{\rm AdS} \rangle = - \frac{2i}{15} \tilde \rho \frac{({\cal K}  - 3 K^{(3)}) \left(  {\cal K} + \frac{3}{2} K^{(3)} \right)}{ {\cal K} + \frac{3}{5} K^{(3)}} .                   
\end{equation}
Combined with the vanishing conditions for the F-terms in the complex structure sector, this allows to write down the stabilisation conditions for the complex structure moduli in terms of their ``dual" periods:
\begin{equation}\label{cpxFterm}
{\cal G}_Q \frac{\hat \rho_K}{\IM({\cal C F}_K)} = - {\cal G}_Q \frac{\hat \rho^\Lambda}{\IM({\cal C Z}^\Lambda)} =    \frac{1}{15}  \tilde \rho\,  \frac{({\cal K}  - 3 K^{(3)}) \left(  {\cal K} + \frac{3}{2} K^{(3)} \right)}{ {\cal K} + \frac{3}{5} K^{(3)}} . 
\end{equation}
Again, these geometric moduli are directly affected by the cubic correction term $K^{(3)}$, in sharp contrast with the axionic moduli.

\section{Conclusions}\label{sec:con}

In this paper we have analysed type IIA orientifold flux vacua taking into account the effect of perturbative $\alpha'$-corrections in the K\"ahler sector. Such corrections are relevant in the sense that they allow to combine the set of RR and NSNS fluxes used to stabilise K\"ahler and complex structure moduli in standard type IIA flux compactifications \cite{Louis:2002ny,Kachru:2004jr,Grimm:2004ua,DeWolfe:2005uu,Camara:2005dc} with an underlying Calabi-Yau geometry \cite{Palti:2008mg}. Such a geometry not only allows to construct a large number of explicit examples, but also simplifies the computation of the 4d effective K\"ahler potential. As a result, one has a large ensemble of flux configurations that can be analysed as a whole. 

Such a set of examples was instrumental in \cite{Herraez:2018vae} to rewrite the F-term scalar potential as a bilinear in flux-axion polynomials $V = Z^{AB}\rho_A\rho_B$. Even if the Calabi-Yau condition is not essential for this reformulation (it can also be obtained for, e.g., twisted tori) it provides  explicit expressions for $Z^{AB}$  in terms of the saxions and for $\rho_A$ in terms of the axions of the compactification. As we have shown, both the bilinear structure and the separate dependence into axions and saxions is maintained in the presence of perturbative $\alpha'$-corrections. This constitutes a proof of concept that the bilinear form of the scalar potential is valid for a large set of flux vacua. It also supports the idea that the factorised dependence into axions and saxions should occur as long as it is a good approximation to assume that fluxes do not affect the 4d K\"ahler metrics of the light fields, or in other words that $Z^{AB}$ is independent of the $\rho_A$. 

We have seen that certain $\alpha'$ corrections modify the definition of the flux-axion polynomials $\rho_A$, in the sense that they redefine the basis of quantised fluxes. Others, namely the cubic correction $K^{(3)}$ that enters the K\"ahler potential as in \eqref{Eq:alphaCorrectedKaehlerPot}, only affect the expression for $Z^{AB}$. Armed with the explicit expressions for both quantities, we have written down the full scalar potential and analysed several of its vacua. We have first considered the class of Minkowski vacua studied in \cite{Palti:2008mg}, and shown that in this case the potential can be written as a bilinear positive definite form \eqref{Eq:ScPotISDAxPol}, as expected from mirror symmetry. The vanishing of each of the entries of the vector \eqref{rhoISD} gives the vacuum conditions for this class of compactifications, and reproduces the results in \cite{Palti:2008mg}. Second, we have considered how $\alpha'$-corrections modify the vacuum conditions of supersymmetric AdS flux vacua, following the same strategy as in \cite{Escobar:2018tiu} and rewriting the vanishing F-term conditions in terms of axion polynomials and solving for them. As in the case of Minkowski vacua, we have found that the cubic correction $K^{(3)}$ only affects the stabilisation of geometric of saxionic moduli, while the other two corrections also affect (implicitly) the stabilisation of axions. 

It would be interesting to extend our results to include more general classes of type IIA flux vacua. For instance one could add open string sectors, like e.g. mobile D6-branes, and see how $\alpha'$-corrections modify the scalar potential in \cite{Herraez:2018vae} and the corresponding vacua analysed in \cite{Escobar:2018tiu}. It would also be interesting to see how the effect of $\alpha'$-corrections modifies the spectrum of soft masses in non-supersymmetric flux vacua, extending the analysis of \cite{Escobar:2018tiu}. In addition, it would also be interesting to compute the effect of perturbative $\alpha'$-corrections for non-Calabi-Yau geometries. In general, we expect that a better understanding of $\alpha'$-corrections in all these cases will allow to root the landscape of type IIA flux vacua on firmer ground. 

\newpage

\centerline{\bf \large Acknowledgments}

\bigskip
We would like to thank I\~naki Garc\'ia-Etxebarria and Eran Palti for useful discussions.  This work is supported by the Spanish Research Agency (Agencia Estatal de Investigaci\'on) through the grant IFT Centro de Excelencia Severo Ochoa SEV-2016-0597, by the grant FPA2015-65480-P from MINECO/FEDER EU, by the grant IJCI-2015-24908 from MINECO, and by the ERC Advanced Grant SPLE under contract ERC-2012-ADG-20120216-320421. D.E. is supported through the FPI grant SVP-2014-068283.


\appendix



\section{Full Computation of the $\alpha'$-Corrected Potentials}\label{A:Potential}
\subsection{$\alpha'$-Corrected K\"ahler Potentials}\label{Aa:ModifiedK}
The compactification of Type IIA string theory on six-dimensional Calabi-Yau orientifolds results in a four-dimensional effective field theory corresponding to ${\cal N}=1$ supergravity coupled to chiral and vector superfields. The chiral supermultiplets comprise of the K\"ahler and complex structure moduli, which parameterise the moduli space associated to the Calabi-Yau metric. In the absence of D-branes, the closed string moduli space factorises properly into a K\"ahler moduli space and a complex structure moduli space, which are both equipped with a moduli-dependent K\"ahler metric arising from the K\"ahler potential:
\begin{equation}\label{Eq:FactorKaehlerPot}
K = K_T + K_Q = - \log({\cal G}_T {\cal G}_Q^2),
\end{equation}
where the functions ${\cal G}_T$ and ${\cal G}_Q$ have been introduced in section~\ref{Ss:TypeIIAFluxVac}. The factorability of the moduli space naturally exhibits itself in the K\"ahler potential as well, yet the product ${\cal G} = {\cal G}_T {\cal G}_Q^2$ can be considered as a whole in which case it is a homogeneous function of degree seven in the geometric moduli $\psi^\alpha \in \{t^a, n^K, u_\Lambda \}$:
\begin{equation}\label{Eq:NoScaleRelHomFunc}
\psi^\alpha \partial_\alpha {\cal G} = \left( t^a \partial_{t^a} + n^K \partial_{n^K} + u_\Lambda \partial_{u_\Lambda}\right) {\cal G} = 7 {\cal G}.
\end{equation}
From this homogeneous function ${\cal G}$, the K\"ahler metric on the moduli spaces can be computed straightforwardly,
\begin{eqnarray}
K_\alpha&=& - \frac{1}{2i} \frac{\partial_\alpha {\cal G}}{{\cal G}}, \\
K_{\alpha \ov \beta} &=&  - \frac{1}{4} \left( \frac{\partial_{\alpha } \partial_\beta {\cal G} }{{\cal G}} - \frac{\partial_\alpha {\cal G} \partial_\beta {\cal G} }{{\cal G}^2} \right). \label{Eq:KaehlMetricHomFunc}
\end{eqnarray}
By employing the homogeneous property~\eqref{Eq:NoScaleRelHomFunc} of the function ${\cal G}$ it is easy to verify that the inverse metric is given by,
\begin{equation}
K^{\alpha \ov \beta} = \frac{2}{3} \psi^\alpha \psi^\beta - 4 {\cal G} {\cal G}^{\alpha \beta},
\end{equation}
where $ {\cal G}^{\alpha \beta}$ is the inverse of $\partial_{\alpha } \partial_\beta {\cal G}$. Property~\eqref{Eq:NoScaleRelHomFunc} also implies some additional relations,
\begin{equation}\label{Eq:ContractKM}
K^{\alpha \ov \beta} K_{\ov \beta} = -2i \psi^\alpha,
\end{equation} 
and 
\begin{equation}
K_\alpha K^{\alpha \ov \beta} K_{\ov \beta} = 7,
\end{equation}
that are typical for no-scale symmetries in the closed string moduli sector of Calabi-Yau (orientifold) compactifications. As pointed out in section~\ref{S:AlphaCorr}, these no-scale symmetries rely on the hidden assumption that we consider large volume regions in the K\"ahler moduli space. Away from this large volume limit, perturbative curvature corrections in $\alpha'$ have to be taken into account, which alter the K\"ahler potential for the K\"ahler moduli space but maintain the factorability of the closed string moduli space. As such, the modified K\"ahler potential derived in~\eqref{Eq:alphaCorrectedKaehlerPot} allows to compute the K\"ahler metric in the same manner as equation~\eqref{Eq:KaehlMetricHomFunc} using the modified function ${\cal G}_T$:
\begin{eqnarray}
K_a &=& \frac{3i}{2} \frac{{\cal K}_a}{{\cal K} \left(1 + \frac{3}{2} \varepsilon \right)}\\
K_{a\ov b} &=& - \frac{3}{2} \frac{1}{{\cal K}^2 \left(1+ \frac{3}{2} \varepsilon \right)^2} \left( {\cal K} (1+ \frac{3}{2} \varepsilon) {\cal K}_{ab} - \frac{3}{2} {\cal K}_a {\cal K}_b  \right),
\end{eqnarray}
while the inverse metric in the presence of perturbative $\alpha'$-corrections is given by:
\begin{equation}\label{Eq:InverseAlphaCorrKM}
K^{a\ov b} =  -\frac{2}{3} {\cal K} \left(1+ \frac{3}{2} \varepsilon \right) \left( {\cal K}^{ab} - 3 \frac{t^a t^b}{{\cal K} (1-3 \varepsilon)} \right).
\end{equation}
Subsequently, relation~\eqref{Eq:ContractKM} is modified as well in the K\"ahler moduli sector due to the curvature corrections:
\begin{equation}\label{Eq:ModifiedRelationKabKb}
K^{a \ov b}K_{\ov b} = - 2i t^a \frac{1+\frac{3}{2} \varepsilon}{1-3 \varepsilon},
\end{equation}
which immediately implies the violation of the no-scale symmetry:
\begin{equation}\label{Eq:ModifiedNoScaleKaehler}
K^{a \ov b} K_a K_{\ov b} = \frac{3}{1-3\varepsilon}.
\end{equation}
This set of relations for the K\"ahler moduli sector turned out to be crucial for the computation of the F-term scalar potential obtained in section~\ref{Ss:AlphaCorrScPot}.

\subsection{$\alpha'$-Corrected Scalar Potential}\label{Aa:ScPotCorrect}
Next, we discuss the computation of the F-term scalar potential in full detail and highlight some manipulations that help us to arrive at the more elegant bilinear form of the scalar potential in equation~\eqref{Eq:ScPotAlphaCorrectBil}. The philosophy used in section~\ref{Ss:AlphaCorrScPot} consists in decomposing the F-term scalar potential in three separate terms and write each term as a function of the ($\alpha'$-corrected) axion polynomials in the simplest form possible. Given that the K\"ahler potentials still factorise between the K\"ahler moduli and complex structure moduli sector, the term containing the derivatives of the superpotential can be written as, 
\begin{equation}
\begin{array}{rcl}
\partial_\alpha W K^{\alpha \ov \beta} \partial_{\ov \beta} \ov W& =&  \partial_{T^a} W K^{T^a \ov T^b} \partial_{\ov T^b} \ov W + \partial_{N^K} W K^{K \ov L}\partial_{N^L} W + \partial_{N^K} W K^{K \ov \Lambda}\partial_{\ov U_\Lambda} W\\
&& + \partial_{U_\Sigma} W K^{\Sigma \ov L}\partial_{N^L} W + \partial_{U_\Sigma} W K^{\Sigma \ov \Lambda}\partial_{\ov U_\Lambda} W\\
&=& K^{a\ov b} (\ov \rho_a - \frac{1}{2} \tilde \rho {\cal K}_a ) (\ov \rho_b - \frac{1}{2} \tilde \rho {\cal K}_b ) + K^{a \ov b} {\cal K}_{ac} \tilde \rho^c {\cal K}_{bd} \tilde \rho^d \\
&&+ K^{N \ov L} \hat \rho_K  \hat \rho_L + K^{N \ov \Lambda} \hat \rho_K  \hat \rho_\Lambda + K^{\Sigma \ov L} \hat \rho_\Sigma  \hat \rho_L + K^{\Sigma \ov \Lambda} \hat \rho_\Sigma  \hat \rho_\Lambda.
\end{array}
\end{equation}
Inserting the expression for the inverse K\"ahler metric~\eqref{Eq:InverseAlphaCorrKM} on the K\"ahler moduli space allows to simplify this relation to the expression in~\eqref{Eq:FtermScPotPart1}. Moreover, this expression can be further rewritten as,
\begin{equation}
\begin{array}{rcl}
\partial_\alpha W K^{\alpha \ov \beta} \partial_{\ov \beta} \ov W& =& K^{a\ov b} \ov \rho_a \ov \rho_b + \frac{4}{9} {\cal K}^2 (1+\frac{3}{2} \varepsilon)^2 K_{a \ov b} \tilde \rho^a \tilde \rho^b + \frac{1 + 6 \varepsilon}{1 - 3\varepsilon} ( {\cal K}_a \tilde \rho^a)^2 + \frac{1}{3} {\cal K}^2 \tilde \rho^2  \frac{(1+\frac{3}{2} \varepsilon)^2}{1-3\varepsilon} \\
&& - \frac{4}{3} \tilde \rho {\cal K} \, \left( \IM W + \frac{1}{6} \tilde \rho {\cal K} + \tilde \rho {\cal K} \varepsilon - \hat \rho_K n^K - \hat \rho^\Lambda u_\Lambda \right)  \frac{(1+\frac{3}{2} \varepsilon)^2}{1-3\varepsilon} \\
&&+ K^{N \ov L} \hat \rho_K  \hat \rho_L + K^{N \ov \Lambda} \hat \rho_K  \hat \rho_\Lambda + K^{\Sigma \ov L} \hat \rho_\Sigma  \hat \rho_L + K^{\Sigma \ov \Lambda} \hat \rho_\Sigma  \hat \rho_\Lambda,
\end{array}
\end{equation}
by eliminating $\ov \rho_a t^a$ through the expression for $\IM W$,
\begin{equation}
\IM W = \ov \rho_a t^a - \frac{1}{6} \tilde \rho {\cal K} - \tilde \rho {\cal K} \varepsilon + \hat \rho_K n^K + \hat \rho^\Lambda u_\Lambda.
\end{equation}
The second component~\eqref{Eq:FtermScPotPart2} is a consequence of imposing the no-scale symmetry in the complex structure moduli sector and the modified relation~\eqref{Eq:ModifiedNoScaleKaehler} in the presence of perturbative $\alpha'$-corrections. The third component~\eqref{Eq:FtermScPotPart3} of the F-term scalar potential results from the factorisation of the K\"ahler potentials~\eqref{Eq:FactorKaehlerPot} for K\"ahler moduli and complex structure moduli, after which one uses the relation~\eqref{Eq:ContractKM} for the complex structure moduli sector and the modified relation~\eqref{Eq:ModifiedRelationKabKb} for the K\"ahler moduli sector. Then, we combine the second component~\eqref{Eq:FtermScPotPart2} and third component~\eqref{Eq:FtermScPotPart3}, upon multiplication by $(1-3\varepsilon)$, to obtain a simplified expression written entirely in terms of the axion polynomials,
\begin{equation}
\hspace{-0.4in}\begin{array}{rcl}
&& (4-3\varepsilon) |W|^2 -4  \left(1+ \frac{3}{2} \varepsilon\right) \left[ \RE W t^a  \partial_{t^a} \RE W +  \IM W t^a \partial_{t^a} \IM W \right]  -4 (1-3\varepsilon) \IM W (n^L \hat \rho_L + u_\Lambda \hat \rho^\Lambda)\\
&&\; = (4-3\varepsilon) \ov\rho_0^2 + 9 {\cal K}_a \tilde \rho^a \ov \rho_0 \varepsilon - ({\cal K}_a \tilde \rho^a)^2 (1 + \frac{15}{4} \varepsilon) - 9\varepsilon\, \IM W
\left( t^a \ov \rho_a -  \hat \rho_K n^K - \hat \rho^\Lambda u_\Lambda  \right) \\
&&\quad  + \tilde \rho {\cal K} \IM W \left( \frac{4}{3} - \frac{1}{2} \varepsilon (1 - 6 \varepsilon)\right).
\end{array}
\end{equation}
Adding up the three components correctly, one can then deduce the final expression for the scalar potential including the perturbative $\alpha'$-corrections, namely equation~\eqref{Eq:FullScPotAlpha}, by manipulating the end result further and separating zeroth order terms ${\cal O}(\varepsilon^0)$, similar to the ones that appear in the inverse metric \eqref{Eq:InverseMetricZNoAlpha}, from higher order $\varepsilon$-corrections.

\subsection{Alternative Computation of the ISD Scalar Potential}\label{Aa:ISDpot}
In section~\ref{Ss:ISDVacua} a positive semi-definite form of the scalar potential in the presence of ISD flux was used to extract the non-supersymmetric Minkowski vacuum configuration in term of the axion polynomials. The precise form of this scalar potential can be derived by a series of computations that start from the T-dual Type IIB picture. In Type IIB the scalar potential associated to ISD flux vacua is explicitly positive definite when expressed in terms of the background ISD $G_3$ fluxes, see e.g.~appendix A of~\cite{Giddings:2001yu},
\begin{equation}\label{Eq:GKPScPot}
\begin{array}{rcl}
V_{GKP}  &=&  \frac{1}{24 V_{{\cal M}_6}} \int_{{\cal M}_6} \frac{|G_3 + i \star_6 G_3|^2}{\IM(\tau)} \\
& =& \frac{1}{2  V_{{\cal M}_6} \IM(\tau) } \int_{{\cal M}_6} (\RE G_3 - \star_6 \IM G_3) \wedge \star_6 (\RE G_3 - \star_6 \IM G_3) ,
\end{array}
\end{equation}
with the three-form flux $G_3 = F_3 - \tau H_3$ and $\tau = C_0 + i\,e^{-\phi}$ the (four-dimensional) axio-dilaton. According the appendix A of~\cite{Palti:2008mg} the Type IIB three-form flux can be T-dualised to the closed string fluxes of Type IIA compactifications, when one considers the following decomposition of the $G_3$-flux in terms of harmonic three-forms,
\begin{equation} 
G_3 = - \left( e_0 + S h_0, e_a, -m,  m^a   \right)  \cdot \left( \begin{array}{c} \beta^0 \\ \beta^a \\ - \alpha_0 \\ -\alpha_a  \end{array}  \right).
\end{equation}
In order to evaluate the scalar potential for the ISD-flux background, one needs to determine the Hodge duals of the harmonic three-forms~\cite{Grimm:2004ua,Palti:2008mg},  
\begin{equation}\label{Eq:HodgeDual}
\star_6 \left( \begin{array}{c} \beta^I \\ - \alpha_I \end{array} \right) = {\cal M}^{-1}  \left( \begin{array}{c} \alpha_I \\ \beta^I \end{array} \right).
\end{equation}
The transformation matrix ${\cal M}^{-1}$ can be further decomposed in terms of the matrices ${\cal R} = \RE({\cal N}_{IJ})$ and ${\cal I} = \IM({\cal N}_{IJ})$, 
\begin{equation}
{\cal M}^{-1} = \left(\begin{array}{cc} {\cal I}^{-1} & - {\cal I}^{-1} {\cal R} \\
-{\cal R} {\cal I}^{-1} & {\cal I} + {\cal R} {\cal I}^{-1} {\cal R}
 \end{array} \right)  = \left( \begin{array}{cc} \mathbb{I}& \\ -{\cal R} &   \mathbb{I}   \end{array} \right)  \left( \begin{array}{cc}{\cal I}^{-1}& \\ & {\cal I}   \end{array} \right)  \left( \begin{array}{cc} \mathbb{I}&-{\cal R}\\ &   \mathbb{I}   \end{array} \right), 
\end{equation}
which follow form the moduli-dependent matrix ${\cal N}_{IJ}$ computed directly~\cite{Grimm:2004ua} from a pre-potential ${\cal F}$,
\begin{equation}
{\cal N}_{IJ} = \ov {\cal F}_{IJ} + 2i\, \frac{ (\IM{\cal F})_{I K} X^K  (\IM{\cal F})_{J L} X^L }{ X^K (\IM{\cal F})_{K L} X^L  },
\end{equation}
where $X^K$ represent the homogeneous coordinates used to parameterise the corresponding moduli space. In the absence of perturbative $\alpha'$-corrections one can insert  the tree-level pre-potential~\eqref{Eq:PrePotTree} for the K\"ahler moduli sector to obtain the respective matrices, while the inclusion of the perturbative curvature corrections requires us to use the modified pre-potential~\eqref{Eq:alphaCorrPrePot}. In the latter case, the resulting transformation matrix ${\cal M}^{-1}$ can be decomposed as
\begin{equation} 
{\cal M}^{-1} = - \frac{3}{2}\mathbb{Q}^t \cdot \mathbb{R}^{-1} \cdot \mathbb{G}^{-1} \cdot \mathbb{R}^{-1t} \cdot \mathbb{Q},
\end{equation}
with $\mathbb{R}$ the axion rotation matrix,
\begin{equation}
\mathbb{R} = \left( \begin{array} {cccc} 1 & & & \\
-b^i & \delta^i_j & & \\
 \frac{1}{6} {\cal K}_{ijk} b^i b^j b^k & -\frac{1}{2} {\cal K}_{ijk} b^j b^k &  1 &  b^i\\
\frac{1}{2} {\cal K}_{ijk} b^j b^k & - {\cal K}_{ijk} b^k & & \delta^i_j  
  \end{array}\right)
\end{equation}
and the lower order curvature corrections $K_{ab}^{(1)}$ and $K_a^{(2)}$ encoded in the matrix $\mathbb{Q}$,
\begin{equation}
\mathbb{Q} =  \left(\begin{array}{cccc}  
1 & 0 & 0& - K_a^{(2)}\\
0 & \delta^a_b & - K^{(2)}_b& -K^{(1)}_{ab}\\
0&  0 & \delta^a_b & 0 \\
0& 0&0&1 
\end{array}
 \right).
\end{equation}
Furthermore, the symmetric matrix $\mathbb{G}^{-1}$ incorporates the curvature corrections proportional to $K^{(3)}$ in the form of the parameter $\varepsilon$,
\begin{equation}
\begin{array}{rcl}
\mathbb{G}^{-1} &=& \frac{1}{{\cal K} (1 +\frac{3}{2} \varepsilon)} \left[ \left( \begin{array}{cccc} 4 & 0 & 0& 0 \\
0 &  K^{a\ov b} & 0& 0\\
0& 0& \frac{{\cal K}^2}{9} \left(1+ \frac{3}{2} \varepsilon \right)^2 & 0\\
0&0&0& \frac{4}{9}  {\cal K}^2 \left( 1+ \frac{3}{2} \varepsilon \right)^2 K_{a\ov b}
 \end{array}\right) \right.\\
 && \qquad \qquad \left.  + \frac{\varepsilon}{1 - 3 \varepsilon} \left( \begin{array}{cccc} 
9& 0& 0& \frac{9}{2} {\cal K}_a \\
 0& - 9 t^a t^b & \frac{3}{2} {\cal K} t^a (1-3\varepsilon) & 0 \\
 0&  \frac{3}{2} {\cal K} t^b (1-3\varepsilon) & - \frac{{\cal K}^2}{4} (1-3 \varepsilon)^2 & 0 \\
 \frac{9}{2} {\cal K}_b & 0 & 0 & \frac{9}{4} {\cal K}_a {\cal K}_b
  \end{array} \right)  \right]
 \end{array}
\end{equation}
By using the Hodge duality relations~\eqref{Eq:HodgeDual} we can rewrite the ISD three-form flux in terms of the matrices $\mathbb{R}$ and $\mathbb{Q}$ and the (modified) axion polynomials~\eqref{Eq:CSgaugeinvariantrhoAlphaCor} as follows,
\begin{equation}
\begin{array}{rcl}
- \RE(G_3) + \star_6 \IM (G_3) &=& 
\left( \begin{array}{c} \ov\rho_0 \\ 
 \ov \rho_a +  \frac{27\varepsilon}{4(1-3\varepsilon)(1+\frac{3}{2} \varepsilon)} \frac{{\cal K}_a}{{\cal K}}  s \hat \rho_0   \\
   -\tilde \rho - \frac{6(1-\frac{3}{4} \varepsilon) }{ (1-3\varepsilon)(1+\frac{3}{2} \varepsilon)} \frac{s}{{\cal K}} \hat \rho_0 \\
    \tilde \rho^a  \end{array} \right)^t \cdot \left(   \mathbb{R} \cdot \mathbb{Q}^{-1 t} \right) \cdot  \left( \begin{array}{c} \beta^0 \\ \beta^a \\ - \alpha_0 \\ -  \alpha_a  \end{array} \right).
\end{array}
\end{equation}
Next, we evaluate the expression of the scalar potential~\eqref{Eq:GKPScPot} for this flux background and use the Hodge duality relations for the harmonic three-forms~\eqref{Eq:HodgeDual}, such that a bilinear structure in terms of the axion polynomials emerges explicitly. After the appropriate Weyl rescaling to 4d Einstein frame we obtain 
\begin{equation}
V_{GKP} = e^{K_T + K_Q} \left( \begin{array}{c} \ov\rho_0 \\ 
 \ov \rho_a +  \frac{27\varepsilon}{4(1-3\varepsilon)(1+\frac{3}{2} \varepsilon)} \frac{{\cal K}_a}{{\cal K}}  s \hat \rho_0   \\
   -\tilde \rho - \frac{6(1-\frac{3}{4} \varepsilon) }{ (1-3\varepsilon)(1+\frac{3}{2} \varepsilon)} \frac{s}{{\cal K}} \hat \rho_0 \\
    \tilde \rho^a  \end{array} \right)^t \cdot  {\cal G}^{-1}  \cdot \left( \begin{array}{c} \ov\rho_0 \\ 
 \ov \rho_a +  \frac{27\varepsilon}{4(1-3\varepsilon)(1+\frac{3}{2} \varepsilon)} \frac{{\cal K}_a}{{\cal K}}  s \hat \rho_0   \\
   -\tilde \rho - \frac{6(1-\frac{3}{4} \varepsilon) }{ (1-3\varepsilon)(1+\frac{3}{2} \varepsilon)} \frac{s}{{\cal K}} \hat \rho_0 \\
    \tilde \rho^a  \end{array} \right).
\end{equation}
with ${\cal G}^{-1} = {\cal K} (1 +\frac{3}{2} \varepsilon) \mathbb{G}^{-1}$.
Note that this expression is equivalent to equation~\eqref{Eq:ScPotISDAxPol} upon rotation of the axion basis by the transformation matrix $\mathbb{T}$, 
\begin{equation}
\mathbb{T} =  \left( \begin{array}{ccccc} 1& 0 & 0& 0 \\
0 & \delta_a^b& 0& 0\\
0&0&0 & -1 \\
0& 0&\delta_b^a & 0 
 \end{array}\right),
\end{equation}
which equally allows to switch between the flux quanta basis $(e_0,e_a,-m,m^a)$ and $(e_0,e_a,m^a,m)$.


\bibliographystyle{JHEP2015}
\bibliography{refs_BilSUSYDFSZ}

\end{document}